\documentclass{aa}
\usepackage{graphicx}
\usepackage{subcaption}
\usepackage{enumitem}
\usepackage[dvipsnames]{xcolor}
\usepackage{amsmath}
\usepackage{txfonts}
\usepackage[colorlinks=True,pdfborder={0 0 0}, linkcolor=blue,citecolor=blue]{hyperref}
\begin{document}

   \title{The birth and early evolution of a low-mass protostar}

   \author{A. Ahmad
          \inst{1}
          \and
          M. González\inst{1}
          \and
          P. Hennebelle\inst{2}
          \and
          B. Commerçon\inst{3}
          }

   \institute{Université Paris Cité, Université Paris-Saclay, CEA, CNRS, AIM, F-91191, Gif-sur-Yvette, France
             \and
             Université Paris-Saclay, Université Paris Cité, CEA, CNRS, AIM, 91191, Gif-sur-Yvette, France
             \and
             Univ Lyon, Ens de Lyon, Univ Lyon 1, CNRS, Centre de Recherche Astrophysique de Lyon UMR5574, 69007, Lyon, France
             }

   \date{Received XXXX; accepted XXXX}

 
  \abstract
  {Understanding the collapse of dense molecular cloud cores to stellar densities and the subsequent evolution of the protostar is of importance to model the feedback effects such an object has on its surrounding environment, as well as describing the conditions with which it enters the stellar evolutionary track. This process is fundamentally multi-scale, both in density and in spatial extent, and requires the inclusion of complex physical processes such as self-gravity, turbulence, radiative transfer, and magnetic fields. As such, it necessitates the use of robust numerical simulations.} 
   {We aim to model the birth and early evolution of a low-mass protostar. We also seek to describe the interior structure of the protostar and the radiative behavior of its accretion shock front.}
   {We carried out a high resolution numerical simulation of the collapse of a gravitationally unstable $1$ $\mathrm{M_{\odot}}$ dense molecular cloud core to stellar densities using 3D radiation hydrodynamics under the gray flux-limited diffusion approximation. We followed the initial isothermal phase, the first adiabatic contraction, the second gravitational collapse triggered by the dissociation of $\mathrm{H}_{2}$ molecules, and $\approx 247$ days of the subsequent main accretion phase.}
   {We find that the subcritical radiative behavior of the protostar's shock front causes it to swell as it accretes matter. We also find that the protostar is turbulent from the moment of its inception despite its radiative stability. This turbulence causes significant entropy mixing inside the protostar, which regulates the swelling. Furthermore, we find that the protostar is not fully ionized at birth, but the relative amount of ionized material within it increases as it accretes matter from its surroundings. Finally, we report in the appendix the results of the first 3D calculations involving a frequency-dependent treatment of radiative transfer, which has not produced any major differences with its gray counterpart.}
   {}

   \keywords{Stars: Formation - Stars: Protostars - Stars: Low-mass - Methods: Numerical - Hydrodynamics - Radiative transfer - Gravitation - Turbulence}

   \maketitle
%

\section{Introduction}

   Despite its common occurrence in the Universe, understanding the collapse of gravitationally unstable dense molecular cloud cores, mostly composed of hydrogen and helium, to stellar densities is a challenging task to overcome in stellar formation theory. This does indeed entail both complex physics and observational challenges that have so far proved extremely difficult to tackle. Newly formed protostellar cores have a typical radius of about $\sim 2$ $\rm{R_{\odot}}$ and are deeply embedded in their opaque parent molecular cloud core. When coupled with the fact that most stars form in regions of our galaxy situated at $\sim$ 100 $\rm{pc}$ within relatively short timescales, observational breakthroughs have been sparse (e.g., \citealp{andre_1993, maury_2019}, see additionally the review by \citealp{dunham_2014}). From a theoretical standpoint, the challenge arises from the complex interplay between numerous physical processes: self-gravitating hydrodynamics, magnetic fields, radiative transfer, and turbulence. In addition, phase transitions such as molecular hydrogen dissociation also need to be taken into account. As a result, an analytical description of protostellar birth is virtually impossible and the field is dominated by numerical models. 
   \\ 
   \\
   The first of such works was that done by \cite{larson1969}, who computed the collapse of a dense molecular cloud core to stellar densities in 1D spherical symmetry. In this pioneering work, Larson identified a two stage evolutionary sequence resulting in the birth of a low-mass protostar. Initially, as the cloud core collapses, any compressive heating generated by the gravitational contraction is immediately radiated away in the infrared by dust grains. This initial isothermal phase is followed by an adiabatic heating phase after the gas density reaches $\sim 10^{-13}$ $\rm{g\text{ }cm^{-3}}$, where the optical depth exceeds unity and radiative cooling becomes inefficient. As a result, the central regions build enough thermal pressure support to reach a state of hydrostatic equilibrium: this is the birth of the first Larson core. It continues its contraction adiabatically with a polytropic index $\gamma_{\mathrm{eff}}$ of five-thirds, which then changes to seven-fifths once temperatures exceed 85 $\rm{K}$ and the rotational degrees of freedom of H$_2$ are excited.
   \\
   Once the temperature of the first Larson core exceeds $2000$ $\rm{K}$, the thermal dissociation of H$_{2}$ is triggered, which is a highly endothermic process that consumes 4.48 $\rm{eV}$ per molecule \citep{stahler2004}. As a result, the energy provided by the compressive heating is mostly spent on the dissociation process instead of providing additional thermal pressure support. This breaks the state of hydrostatic equilibrium, and a violent second collapse ensues with $\gamma_{\mathrm{eff}}$ $\approx 1.1$. The extreme rise in density and temperature following this event gives birth to a new protostellar object in hydrostatic equilibrium: the second Larson core\footnote{We sometimes refer to this object as the protostar.}. The protostar continues accreting material from the infalling envelope, and angular momentum conservation leads to the formation of a circumstellar disk. Once core temperatures exceed $10^{6}$ $\rm{K}$, deuterium burning begins, thus ending the pre-stellar phase.
   \\
   \\
   This evolutionary sequence has so far been well accepted for low-mass protostars. Since the work done by \cite{larson1969}, the field has developed ever more robust codes to tackle the 21 orders of magnitude in density and eight in spatial extent, in fully 3D simulations in order to include the effects of magnetic fields, rotation, turbulence, as well as radiation (for a detailed summary of each milestone reached over the years, see \citealp{teyssier_2019}). These advancements were brought about by the ever increasing amount of computing power available. However, this growing complexity of the simulations has also meant that their computational costs has increased. As a result, there is a vast parameter space to explore and determine the role different physical processes play, but this task is hindered by the technical costs of the simulations. Such technical difficulties have significantly constrained the time stepping in self-consistent 3D simulations, and current state-of-the-art papers struggle to integrate the calculations past a few years after the birth of the protostar (e.g., \cite{vaytet_2018} reached 24 days using adaptive mesh refinement, \cite{wurster_2020} reached 4 years using smooth particle hydrodynamics, whereas the 1D code in \cite{masunaga_2000} reached $1.3\times 10^{5}$ years). Such constraints have forced researchers interested in larger timescales to omit the expensive calculations of the protostar by replacing it with a sink particle \citep{bate_1995, bleuler_2014}, effectively reducing the feedback effects the protostar has on larger spatial scales to a sub-grid model (e.g., \citealp{vorobyov_2015, tomida_2017, hennebelle_disks, wurster_disks, lebreuilly_2021}).
   \\
   \\
   Despite the many advancements achieved over the years, the difficulties in integrating the simulations across large timescales has meant that the evolution of the protostar is still poorly understood. Since protostellar feedback plays a significant role in the formation and fragmentation of its surrounding disk, the temperature and structure of its envelope, as well as the overall dynamics of molecular clouds \citep{hennebelle_2020, hennebelle_2022, grudic_2022}, understanding the physics at the protostellar scale is of crucial importance.
   Hence, our goal is to model the birth of the protostar and study its evolution through time in a self-consistent 3D manner. We place a special focus on the interior structure of the protostar, its accretion shock, and the inner turbulent motions, in order to understand its behavior. Since previous studies in the literature involving nonideal magneto-hydrodynamics have shown that protostars are born with weak magnetic field strengths, ranging from $10^{-1}-10^{3}\ \mathrm{G}$ \citep{vaytet_2018, wurster_2020, wurster_2022}, the thermal pressure is orders of magnitude above the magnetic pressure. Hence, we have decided to omit magnetic fields from our study and have constrained ourselves to a radiation-hydrodynamics (RHD) model under the gray flux-limited diffusion (FLD) approximation. This provides the added benefit of reducing the computational cost of the simulations. Our simulations were carried out using the adaptive mesh refinement (AMR) code {\ttfamily RAMSES} \citep{teyssier_2002}. In addition, we have for the first time carried out a 3D simulation with frequency-dependent radiative transfer leading to the formation of the protostar. Its results are in agreement with its gray counterpart, and we have reported them in the appendix.
   \\
   In \hyperref[section:model]{Sec. \ref*{section:model}}, we present the numerical methods and the initial conditions used in this work. The birth of the protostar, its evolution through time, and its chemical composition are presented in \hyperref[section:results]{Sec. \ref*{section:results}}. Finally, the behavior of the turbulence found within the protostar is studied in \hyperref[sec:mainturb]{Sec. \ref*{sec:mainturb}}.

\section{Model} \label{section:model}
\subsection{{\ttfamily RAMSES} with multigroup flux limited diffusion}
Our simulations were carried out using the 3D adaptive mesh refinement and finite-volume code {\ttfamily RAMSES} \citep{teyssier_2002}. In order to include radiative transfer, we have used the flux-limited diffusion module developed by \cite{commercon_2011, commercon_2014}, and its extension to a multigroup description by \cite{gonzalez_2015}. Since the protostar and its evolution over time are our subject of interest, we have naturally chosen the gray (single-group) approximation, which allows for better performance. However, we have also run a simulation with a multigroup description in order to compare it with its gray counterpart, the results of which are presented in \hyperref[appendix:MultiGroup]{Appendix \ref{appendix:MultiGroup}}. Hence, for the sake of clarity, we present our governing equations in their general (multigroup) form, which consist of the Euler equations coupled with a radiative energy equation \citep{gonzalez_2015}:
\\
\begin{equation}
\label{eq:continuity}
    \frac{\partial\rho}{\partial t}+\Vec{\nabla}\cdot [\rho \Vec{\mathrm{v}}] = 0\ ,
\end{equation}
\begin{equation}
\label{eq:conservmom}
    \frac{\partial\rho \Vec{\mathrm{v}}}{\partial t} + \Vec{\nabla}\cdot \left [ \rho\Vec{\mathrm{v}} \otimes \Vec{\mathrm{v}} + P\mathbb{I} \right ] = -\rho\Vec{\nabla}\phi - \sum_{g=1}^{N_{g}}\lambda_{g}\Vec{\nabla}E_{g}\ ,
\end{equation}
\begin{multline}
\label{eq:conservener}
    \frac{\partial E_{\mathrm{tot}}}{\partial t} + \Vec{\nabla}\cdot \left [ \Vec{\mathrm{v}}(E_{\mathrm{tot}} + P) \right ] = -\rho\Vec{\mathrm{v}}\cdot\Vec{\nabla}\phi \\- \sum_{g=1}^{N_{g}}\left [\kappa_{\mathrm{P}_g}\rho c (\Theta_{g}(T)-E_{g}) - \lambda_g\Vec{\mathrm{v}}\cdot\Vec{\nabla} E_{g} \right ]\ ,
\end{multline}
\begin{multline}
\label{eq:conservenerEr}
    \frac{\partial E_{g}}{\partial t} + \Vec{\nabla}\cdot [\Vec{\mathrm{v}}E_{g}] + \mathbb{P}_{g} : \Vec{\nabla}\Vec{\mathrm{v}} = \Vec{\nabla} \cdot \left[ \frac{c\lambda_g}{\rho \kappa_{\mathrm{R}g}} \Vec{\nabla} E_g\right] \\ +\kappa_{\mathrm{P}g} \rho c \left(\Theta_g(T)-E_g\right) + \Vec{\nabla} \Vec{\mathrm{v}} : \int_{\nu_{g-1/2}}^{\nu_{g+1/2}}{\partial_\nu(\nu \mathbb{P}_\nu) d\nu}\ ,
\end{multline}
\begin{equation}
\label{eq:poisson}
    \nabla^{2}\phi = 4\pi G \rho\ ,
\end{equation}
where $\rho$ is the gas density, $\Vec{\mathrm{v}}$ its velocity vector, $P$ its thermal pressure, $T$ its temperature, $\phi$ the gravitational potential, $\mathbb{I}$ the identity operator, $\kappa_{\mathrm{P}g}$ the Planck mean opacity, $\kappa_{\mathrm{R}g}$ the Rosseland mean opacity, $G$ the gravitational constant, $c$ the speed of light, and $\lambda_g$ the flux limiter. We note that $N_{g}$ is the total number of radiative groups whose frequency borders are $\nu_{g\pm 1/2}$. $\Theta_{g}$ is the energy carried by photons that have a Planck distribution of temperature $T$ inside their given radiative group. $E_{\mathrm{tot}}$ is the total gas energy, which includes the kinetic and internal energy $E$:
\begin{equation}
    E_{\mathrm{tot}} = \frac{1}{2} \rho v^2 + E\ .
\end{equation}
$E_g$ (resp. $\mathbb{P}_{g}$) is the frequency-integrated radiative energy (resp. pressure tensor) inside each group:
\begin{equation}
    E_{g} = \int_{\nu_{g-1/2}}^{\nu_{g+1/2}} E_{\nu} d\nu\ , \\
    \mathbb{P}_{g} = \int_{\nu_{g-1/2}}^{\nu_{g+1/2}}\mathbb{P}_{\nu} d\nu\ .
\end{equation}
The opacities are also computed in the same manner:
\begin{equation}
    \kappa_{\mathrm{R}g} = \int_{\nu_{g-1/2}}^{\nu_{g+1/2}} \kappa_{\mathrm{R}\nu} d\nu\ , \\
    \kappa_{\mathrm{P}g} = \int_{\nu_{g-1/2}}^{\nu_{g+1/2}} \kappa_{\mathrm{P}\nu} d\nu\ .
\end{equation}
We define the total radiative energy $E_{\mathrm{r}}$ as the sum of the radiative energy inside each group:
\begin{equation}
    E_{\mathrm{r}} = \sum_{g=1}^{N_{g}} E_{g}\ .
\end{equation}
\\
\\
\hyperref[eq:continuity]{Equation \ref*{eq:continuity}} is the continuity equation, \hyperref[eq:conservmom]{Eq. \ref*{eq:conservmom}} describes the conservation of momentum, \hyperref[eq:conservener]{Eq. \ref*{eq:conservener}} the conservation of energy, \hyperref[eq:conservenerEr]{Eq. \ref*{eq:conservenerEr}} the conservation of radiative energy, and \hyperref[eq:poisson]{Eq. \ref*{eq:poisson}} the Poisson equation for self-gravity.
\\
\\
The code uses the HLL Riemann solver to solve the hydro equations, and the radiative energy equations are solved using a time implicit solver with the following flux limiter \citep{minerbo_1978}:
\begin{equation}
      \lambda_{g} = \begin{cases}
    2/\left ( 3+\sqrt{9+12R_{g}^{2}}\right )\ , & \text{if $0$ $\leq$ $R_{g}$ $\leq 3/2$}\\
    \left ( 1+R_{g}+\sqrt{1+2R_{g}}\right )^{-1}, & \text{if $3/2 <$ $R_{g}$ $\leq\infty$}
  \end{cases}
\end{equation}
with $R_{g} = |\Vec{\nabla}E_{g}|/(\rho\kappa_{\mathrm{R}_{g}}E_{g})$. The radiative pressure tensor is given by:
\begin{equation}
    \mathbb{P}_{g} = \left ( \frac{1-\chi_{g}}{2}\mathbb{I} + \frac{3\chi_{g}-1}{2}\Vec{n_{g}}\otimes\Vec{n_{g}} \right ) E_{g}\ ,
\end{equation}
where $\chi_{g} = \lambda_{g}+\lambda_{g}^{2}R_{g}^{2}$ and $\Vec{n_{g}} = \Vec{\nabla}E_{g}/|\Vec{\nabla}E_{g}|$. Under the optically thick limit, $R_{g}\to 0$ and $\lambda_{g}\to 1/3$ which causes $\mathbb{P}_{g}$ to become isotropic. In the main body of this paper, we have used the gray approximation, meaning that there is a single group of photons (i.e., $N_{g}=1$).
\\
\\
The equation of state used is the tabulated EOS of \cite{saumon_1995}, which has been extended to lower densities by \cite{vaytet_2013}. It describes the thermal properties of $\mathrm{H_2}$, $\mathrm{H}$, $\mathrm{H^+}$, $\mathrm{He}$, $\mathrm{{He}^{+}}$, and $\mathrm{{He}^{2+}}$. The cloud has an initial mixture of 73\% H and 27\% He.
\\
The gas and dust opacities were taken from \cite{vaytet_2013}, who pieced together a table of opacities in the range of $10^{-19}$ $\rm{g\text{ }cm^{-3}} < \rho < 10^{2} $ $\rm{g\text{ }cm^{-3}}$ and 5 $\rm{K} <T <10^{7}$ $\rm{K}$ from \cite{semenov_2003}, \cite{ferguson_2005} and \cite{badnell_2005} (see Figure 2 of \citealp{vaytet_2013}). When temperatures are below 1500 $\rm{K}$, the dust particles (which represent 1\% of the mass content of the fluid) dominate the opacities and they are in thermal equilibrium with the gas. Once temperatures exceed 1500 K, the dust sublimates and the molecular gas opacities begin to dominate. Finally, when the temperatures exceed 3200 K, all molecules are dissociated and the atomic gas opacities dominate. The Planck and Rosseland mean opacity tables are computed within each frequency group according to the Delaunay triangulation process described in \cite{vaytet_2013}. In gray radiative transfer simulations, there is only a single frequency group ([$10^{5};10^{19}$] $\rm{Hz}$) along which the entire opacities are integrated. The resulting opacity mesh is presented in \hyperref[fig:OPACITYMESH]{Fig. \ref*{fig:OPACITYMESH}}, and the temperature-density distribution of the cells in our computational domain at the epoch of protostellar birth is overlaid in red. At low temperatures, the dust dominates the fluid's opacity; however, they are destroyed once temperatures exceed $\approx 1500$ $\rm{K}$ and the subsequent drop in $\kappa_{\mathrm{R}}$ is clearly visible in the figure. Once the gas transitions toward higher densities, the atomic gas opacities begin to rise and a new opacity peak appears.
\\
\\
Limiting ourselves to a RHD model is not without merit. Indeed, not only does this significantly reduce the computational costs of our simulations, it also has a physical justification. Current state-of-the-art papers involving nonideal MHD have consistently shown that the protostar is born with a weak magnetic field strength, thus placing the magnetic pressure orders of magnitude below the thermal pressure. One can thus omit magnetic fields when describing protostars prior to the beginning of a dynamo process.

\subsection{Initial Conditions}
Our initial conditions consists of a uniform density sphere of mass $M_{0}=$ $1$ $\rm{M_{\odot}}$, initial temperature $T_{0} = 10$ $\rm{K}$, and a radius of $R_{0}=2.465\times10^{3}$ $\rm{AU}$. This molecular cloud core is $100$ times denser than its surrounding environment, and its ratio of thermal to gravitational energies is
\begin{equation} \label{eq:alpha}
    \alpha = \frac{5R_{0}\kappa_{\mathrm{B}}T_{0}}{2GM_{0}\mu m_{\mathrm{H}}} = 0.25\ ,
\end{equation}
where $\kappa_{\mathrm{B}}$ is Boltzmann's constant and $m_{\mathrm{H}}$ is the atomic mass constant. The mean molecular weight $\mu$ corresponds to 2.31 for our initial gas mixture.
\\
\\
As we have chosen to focus our attention on the formation and early evolution of the protostar, we have not included any motion in our initial conditions, be it in the form of coherent solid body rotation or any turbulent velocity vector field in the cloud core. This allows the ensuing gravitational collapse to form a spherical, central protostar in the absence of any disks. Hence, our computational resources are more devoted to the protostar, and we can integrate our simulations for longer timescales. In this respect, our study is equivalent to 1D calculations such as those of \cite{larson1969}, \cite{narita_1970}, \cite{winkler_1980}, \cite{masunaga_2000}, \cite{vaytet_2013}, \cite{vaytet_2017}, \cite{bhandare_2018}, or \cite{bhandare_2020}. The added benefit of carrying out these calculations in 3D is the ability to describe the turbulent motion within the second core, recently brought to light by the 2D study of \cite{bhandare_2020}. As such, these initial conditions provide us with an ideal scenario to study the accretion shock and the interior structure of the protostar.
\subsection{Refinement strategy}
In order to resolve the interior of the protostar, an exceptionally high resolution is required. We continuously refine our AMR grid according to a modified Truelove criterion \citep{truelove_1997}:
\begin{equation}
\label{eq:deltaX}
    \Delta x \leq \frac{\lambda_{\mathrm{j}}^{*}}{N}\ ,
\end{equation}
where $\Delta x$ is the cell length and $N=20$. $\lambda_{\mathrm{j}}^{*}$ is the Jeans length computed at the cell's given density and at a temperature of $100$~$\mathrm{K}$:
\begin{equation}
\label{eq:jeansL}
      \lambda_{\mathrm{j}}^{*} = \begin{cases} 
    \lambda_{\mathrm{j}}\sqrt{\frac{100\text{ }\rm{K}}{T}} & \text{if $T>100$ $\rm{K}$}\\
    \lambda_{\mathrm{j}} & \text{otherwise}
  \end{cases}
\end{equation}
where $\lambda_{\mathrm{j}}$ is the Jean's length. This allows the resolution to follow a length that varies in $\rho^{-1/2}$ independently of temperature once $T>100$ $\rm{K}$. The coarse grid has a resolution of $64^{3}$ cells ($\ell_{\mathrm{min}}=6$), and we allow 20 additional levels of refinement ($\ell_{\mathrm{max}}=26$). This results in an effective spatial resolution of $\Delta x = 1.4\times 10^{-4} \rm{AU}$ at the maximum refinement level. Although some of the protostar's properties are not converged at this resolution (see \hyperref[appendix:ResStudy]{Appendix \ref*{appendix:ResStudy}}), we have nonetheless proceeded with it in order to circumvent the stringent time-stepping constraints.
\\
Our refinement strategy provides us with $N\sqrt{\frac{T}{100\text{ }\rm{K}}}$ cells per actual Jeans length (until the maximum refinement level is reached), which throughout our simulation corresponds to $20-2\times 10^{3}$ cells. In the protostar's central region, we have $\approx 60$ cells per jeans length. This allows us to effectively resolve turbulent motions within the protostar.
\\
\\
Our simulation was run on two nodes, each containing 32 CPU cores. As reported in \cite{vaytet_2018}, the load balancing performs poorly in {\ttfamily RAMSES} when simulating second gravitational collapses, as the majority of the computational load is contained in a small central region. As such, a smaller CPU workforce is the optimal choice as it reduces the MPI communications load. The simulation was run for a total of 2053.75 hours, which corresponds to a usage of 131440 CPU hours. By using the \cite{berthoud_2020} estimate of 4.68 $\mathrm{g}\ \mathrm{hCPU^{-1}}$, the $\mathrm{CO}_{2}$ equivalent carbon footprint of our simulation is $\approx 615\ \mathrm{kg}$.

\begin{figure}
    \centering
    \includegraphics[width=.5\textwidth]{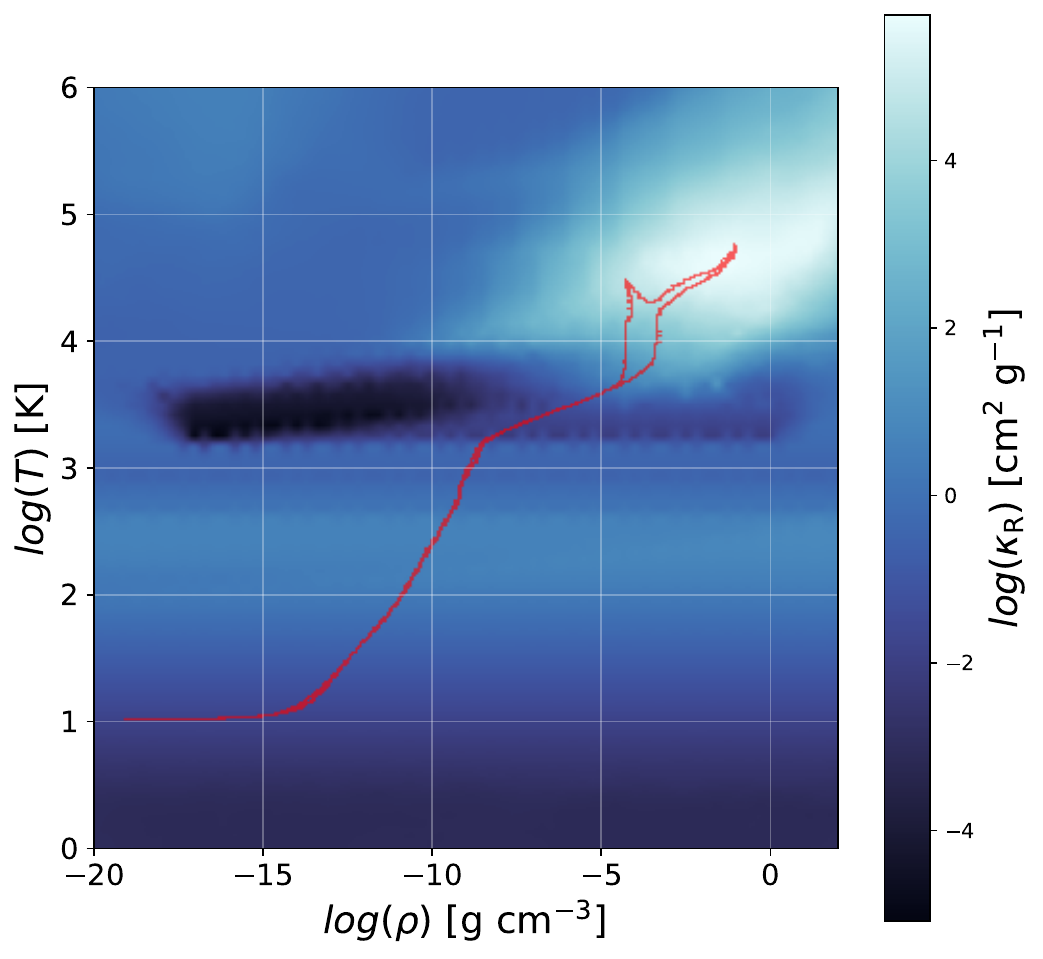}
    \caption{Opacity mesh created for our gray radiative transfer approximation. The temperature-density distribution of all cells during the epoch of protostellar birth is overlaid in red.}
    \label{fig:OPACITYMESH}
\end{figure}

\section{Results} \label{section:results}
\subsection{Genesis}
\begin{figure*}
    \centering
    \includegraphics[width=1\textwidth]{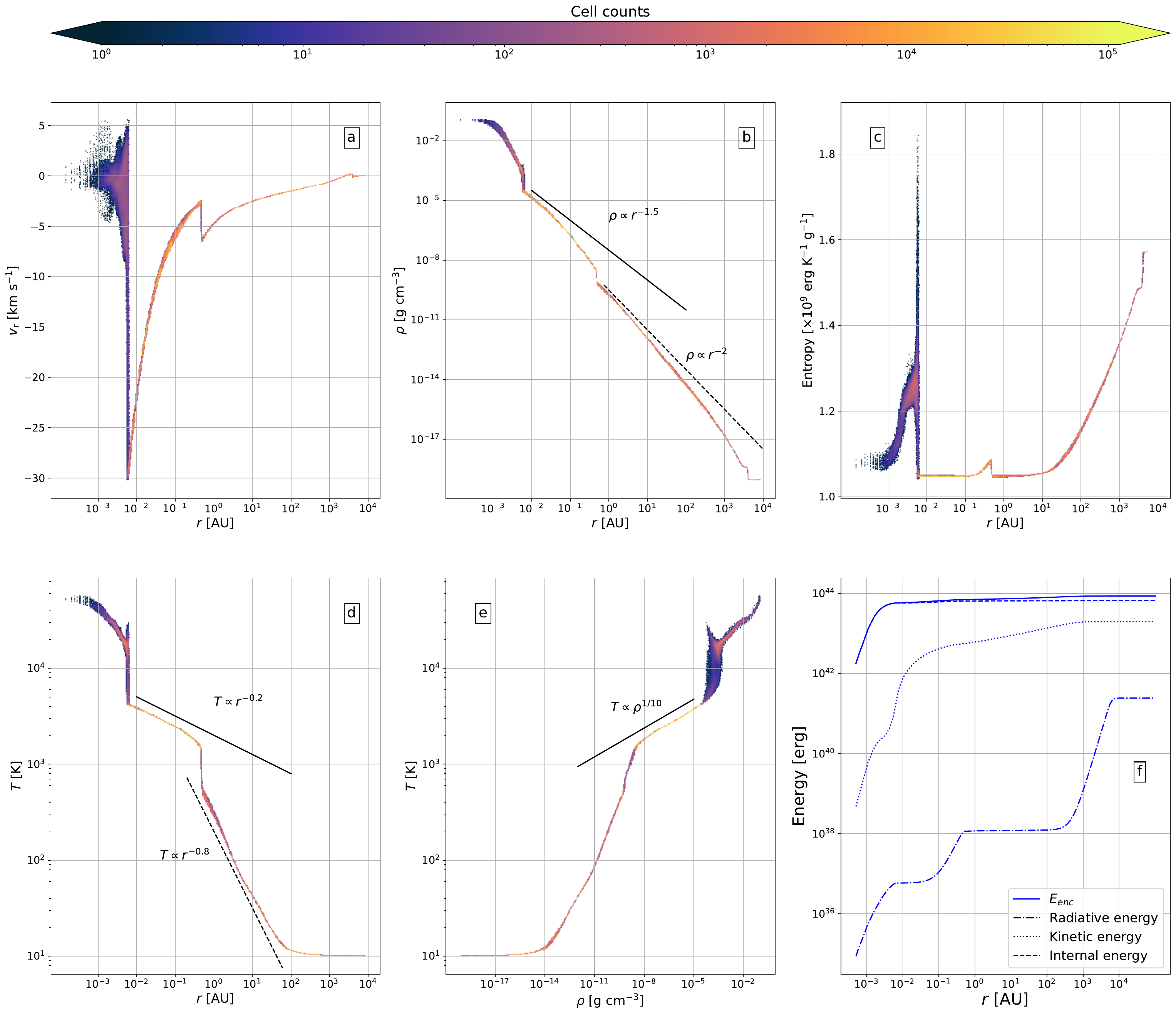}
    \caption{Various sets of 2D histograms binning the cells in our computational domain (panels a-e) at the epoch of protostellar birth. Panels (a), (b), (c), and (d) represent respectively radial velocity, density, entropy, and temperature as a function of radius. The solid (resp. dashed) black line in panel (b) displays the expected density profile for a free-falling gas (resp. for the collapse of an isothermal sphere). The solid (resp. dotted) black line in panel (d) represents the expected temperature profile for the collapse of an isothermal sphere with $\gamma_{\mathrm{eff}}=1.1$ (resp. $\gamma_{\mathrm{eff}}=7/5$). Panel (e) displays temperature as a function of density, where the overlaid solid black line displays a contraction with $\gamma_{\mathrm{eff}}=1.1$. Panel (f) represents the sum of the enclosed gas and radiative energies at radius $r$ (solid line, see \hyperref[eq:enclosedenerg]{Eq. \ref*{eq:enclosedenerg}}), along with its constituent parts, namely internal (dashed line), kinetic (dotted line), and radiative energies (dash-dotted line).}
    \label{fig:genesisHists}
\end{figure*}

\begin{figure*}
    \centering
    \includegraphics[width=.8\textwidth]{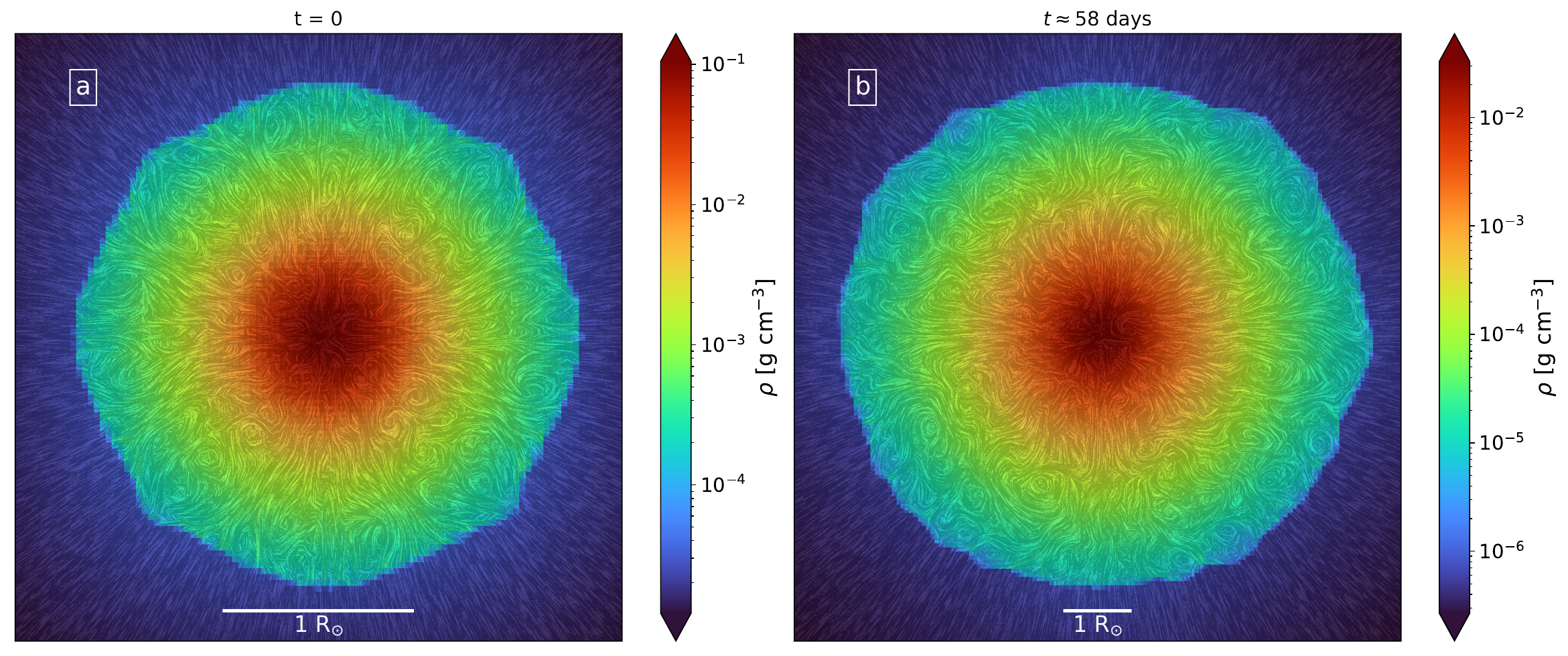}
    \caption{Density slices through the center of the domain at the birth of the protostar (t = 0, panel a) and rougly two months later (t $\approx$ 59 days, panel b). The swirly patterns are line integral convolution (LIC) visualizations of the velocity vector field, which display prominent eddies inside the newly formed protostar. Over the span of $\approx$ two months, the protostar has grown in radius by a factor $\approx$ 2.8.}
    \label{fig:LIC}
\end{figure*}

\begin{figure}
    \includegraphics[width=.45\textwidth]{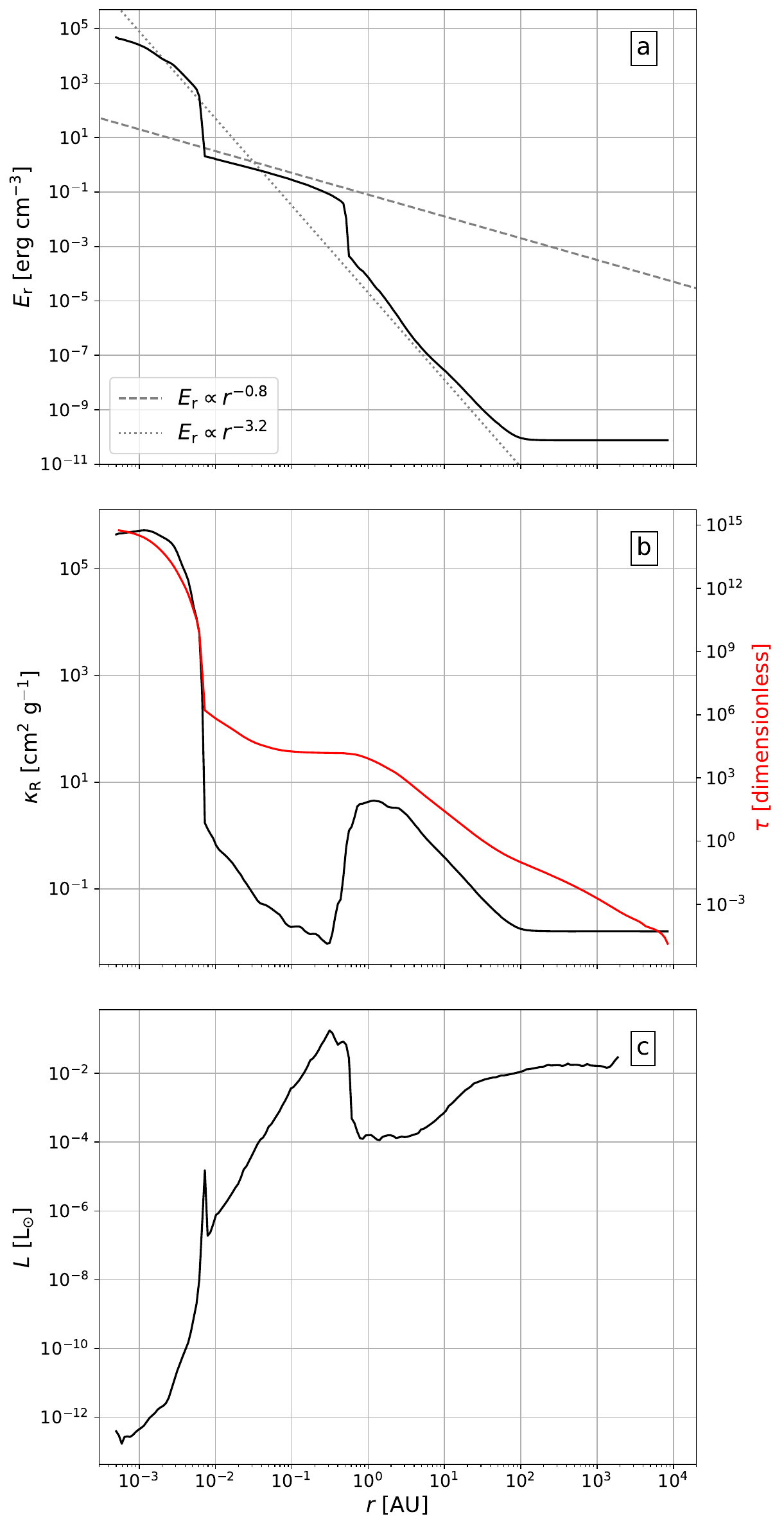}
    \caption{Radiative energy (panel a), Rosseland mean opacity (black, panel b), optical depth (red, panel b), and luminosity (panel c), averaged in radial bins and displayed as a function of radius at the epoch of the protostar's formation.}
    \label{fig:EKL}
\end{figure}

We first begin by describing the system at the epoch of protostellar birth. We define this moment as the instant a second accretion shock forms (i.e., a discontinuity in the radial velocity profile). In \hyperref[fig:genesisHists]{Fig. \ref*{fig:genesisHists}}, we show plots displaying various physical profiles along radius and density. Panel (e) shows the temperature-density distribution of our cells. Here, the previously mentioned two step evolutionary sequence is clearly visible: the collapse begins isothermally, contracts adiabatically, and once the dissociation of $\mathrm{H}_{2}$ begins, a second collapse occurs where $T\propto \rho^{1/10}$ ($\gamma_{\mathrm{eff}}\approx 1.1$). The supersonic free-falling gas then collides with the protostellar surface which causes the shock heating observed after $\rho\sim10^{-5}$ $\rm{g\text{ }cm^{-3}}$, and the gas begins a second phase of adiabatic contraction as the newly formed protostar continues accreting material. The temperatures inside the protostar reach upward of $\approx 8.5\times 10^{4}$ $\rm{K}$. This is a far-cry from the $10^{6}$ $\rm{K}$ needed to fuse deuterium; the protostar must further contract and its core temperature needs to increase ten-fold in order to become a star and join the main sequence.
\\
Panel (a) shows the radial velocity profile, where one can observe a prominent discontinuity at $6\times 10^{-3}$ $\rm{AU}$ ($\approx1.3$ $\rm{R_{\odot}}$), which marks the protostar's border. Another discontinuity, this time corresponding to the first Larson core border, is visible at $0.5$ $\rm{AU}$. The location of these shock fronts also correspond to steep density and temperature gradients in panels (b) and (d). Both outside and inside the first core border, the density profile approaches $\rho\propto r^{-2}$ (dashed black line in panel b), which is characteristic of the collapse of an isothermal sphere \citep{larson1969, penston_1969}. Just outside the second core border, the density profile closely approaches $\rho\propto r^{-1.5}$ (solid black line in panel b), which demonstrates that the accreted gas is free-falling into the newly formed protostar. Since $T\propto \rho^{\gamma_{\mathrm{eff}}-1}$, we also see two differing temperature profiles in panel (d); outside the first core border, the contraction occurs with $\gamma_{\mathrm{eff}}=7/5$, hence $T\propto r^{-0.8}$ (dashed black line). However, inside the first core the contraction occurs with $\gamma_{\mathrm{eff}}\approx 1.1$. As a result, the temperature profile follows $T\propto r^{-0.2}$ (solid black line).
\\
When the free-falling gas reaches the stellar surface, the supersonic collision heats it significantly, as it cannot dissipate its kinetic energy in the form of radiation in these extremely high optical depths (see \hyperref[fig:EKL]{Fig. \ref*{fig:EKL}}, panel b). This causes the temperature spike seen in panel (d) at the second core border, which exceeds the temperature in the protostar's outer layer. This exhibits the radiative nature of the protostar at birth; it mainly radiates the accretion energy it receives at the shock front which far outweighs the cooling flux that struggles to escape the opaque interior. Once inside the protostar, there is a significant amount of spread around $v_{\mathrm{r}}=0$, which shows that there are parcels of fluid that are both rising and falling, thus hinting at the presence of turbulent motions in the protostar's interior. Indeed, when visualizing the velocity vector field in \hyperref[fig:LIC]{Fig. \ref*{fig:LIC}}, there is a significant amount of eddies visible downstream of the accretion shock.
\\
In panel (c), the radial entropy\footnote{The entropy was obtained through an interpolation of the EOS table.} profile is displayed. Here, we once again see two steep gradients corresponding to both core borders. Inside both cores, the entropy profile rises with the radius. This implies that the core is radiatively stable, and cannot generate any convection from its central regions\footnote{This is consistent with the 2D results of \citep{bhandare_2020}.} \citep{stahler2004}. The nature of this turbulent motion will be studied in detail in \hyperref[sec:turb]{Sec. \ref*{sec:turb}}. We subsequently also revisit the behavior of the entropy profile in \hyperref[section:evol]{Sec. \ref*{section:evol}}.
\\
In panel (f), we display the sum of the enclosed gas and radiative energies $E_{\mathrm{enc}}$ as well as its constituent parts, namely radiative, kinetic, and internal energy, as a function of radius, and computed using
\begin{equation}
    \label{eq:enclosedenerg}
    E_{\mathrm{enc}}(r) = 4\pi\int_{0}^{r} \left ( E_{\mathrm{tot}}+E_{\mathrm{r}} \right ) r^{2}dr\ .
\end{equation}
Throughout the entire volume of our computational domain, the bulk of the system's energy resides under internal energy form, and kinetic energy is the second most prominent form. The majority of $E_{\mathrm{enc}}$ is within the protostar itself. By looking at the enclosed radiative energy curve, we can distinguish three plateaus. The first one, just outside the protostar's border, shows that the bulk of the radiative energy at $r<0.1\ \mathrm{AU}$ is located inside the protostar, and is a consequence of the weak radiative energy gradient outside the second core border (see \hyperref[fig:EKL]{Fig. \ref*{fig:EKL}}, panel a). This is also suggesting that very little radiation is escaping the protostar, thus hinting at the subcritical nature of the second core accretion shock (the radiative flux escaping the shock front is inferior to the incoming energy flux). The second plateau, located outside the first core border, is in fact not a real plateau; the enclosed radiative energy is indeed increasing. However there is far too little radiative energy outside the first core to lift the curve any further. Once $r>10^{2}$ $\rm{AU}$, the enclosed radiative energy curve increases once again, as the volume integral now includes the photons emitted by the isothermal phase of the contraction. Finally, the third plateau is simply caused by the fact that we have reached the boundaries of the simulation box, and no new cells are used to compute the volume integral.
\\
\\
We now turn to studying the radiative behavior of the simulation at the birth of the protostar. \hyperref[fig:EKL]{Fig. \ref*{fig:EKL}} shows the specific radiative energy (panel a), and the opacity (black curve in panel b), averaged in radial bins and displayed as a function of radius. The red curve in panel (b) shows the optical depth $\tau$ computed from the outer edge of the simulation box:
\begin{equation}
    \tau = \int_{r}^{3R_{0}} \rho(r)\kappa_{\mathrm{R}}(r)dr.
\end{equation}
\noindent Panel (c) of this figure displays the luminosity $L(r)$, computed as:
\begin{equation}
    \label{eq:lumr}
    L(r) = 4\pi r^{2}c\frac{\lambda(r)\nabla E_{\mathrm{r}}(r)}{\rho (r)\kappa_{\mathrm{R}}(r)}\ .
\end{equation}
Panel (a) shows us that the radiative energy is constant at large radii ($\frac{dE_{\mathrm{r}}}{dr} = 0$). Since the photons being produced locally by the gas are streaming through an optically thin medium, $E_{\mathrm{r}}$ remains constant during this phase of isothermal contraction. Once the gas becomes optically thick to radiation, we witness a subsequent buildup in radiative energy. A sharp gradient, corresponding to the first core accretion shock, is then seen at $0.5$ $\rm{AU}$. It should be noted however that the first core accretion shock has already radiated a substantial amount of energy, which has then propagated outward. This is made possible by the supercritical nature of the first core accretion shock \citep{commercon_shock, vaytet_2018}. Inside the first core, the radiative energy gradient is not as steep as that of the adiabatic gas outside of it. Since $E_{\mathrm{r}}\propto T^{4}$, we have $E_{\mathrm{r}}\propto r^{-3.2}$ outside the first core ($\gamma_{\mathrm{eff}}=7/5$, gray dotted line), whereas $E_{\mathrm{r}}\propto r^{-0.8}$ inside it ($\gamma_{\mathrm{eff}}=1.1$, gray dashed line). The temperatures found inside the first core exceed the dust sublimation temperature ($\approx 1200$ $\rm{K}$), causing the drop in opacity seen in panel (b). Once we reach the protostar, the high densities spike the atomic gas opacities, and the optical depth reaches a staggering $10^{15}$. This causes the steep radiative energy gradient at the protostar's border ($\approx 6\times 10^{-3}$ $\rm{AU}$) and the subsequent buildup seen in its interior.
\\
In the luminosity profile shown in panel (c), we see a spike at the protostar's border. This is the second core accretion shock. Due to the temperature of the shock front, mainly Ultra-Violet photons are emitted at this radius, which are quickly reabsorbed by the optically thick gas upstream and reemitted in the infrared\footnote{The multigroup simulation that we have run and presented in \hyperref[appendix:MultiGroup]{Appendix \ref*{appendix:MultiGroup}} permits us to better distinguish what photon frequencies are produced at all radii.}. As such, the total luminosity exiting the protostellar surface should be measured just upstream of the shock front, which yields a value of $\approx$ $8\times10^{-7}$ $\rm{L_{\odot}}$. Curiously, the total luminosity becomes somewhat constant with the radius starting at $20$ $\rm{AU}$, which shows that the emanating radiative flux decreases as $F_{\mathrm{rad}} \propto r^{-2}$. This means that the photosphere of the system is located at about this radius. The salient question one might ask here is how the system's behavior within the photosphere impacts the amount of flux escaping it, as that would allow us to link our current theoretical understanding of newly formed protostars with photometric observations. However, we have not been able to integrate our calculations long enough to witness any noticeable change in the radiative behavior of the photosphere.

\subsection{Evolution of the protostar}
\label{section:evol}
\begin{figure*}
    \centering
    \includegraphics[width=1.\textwidth]{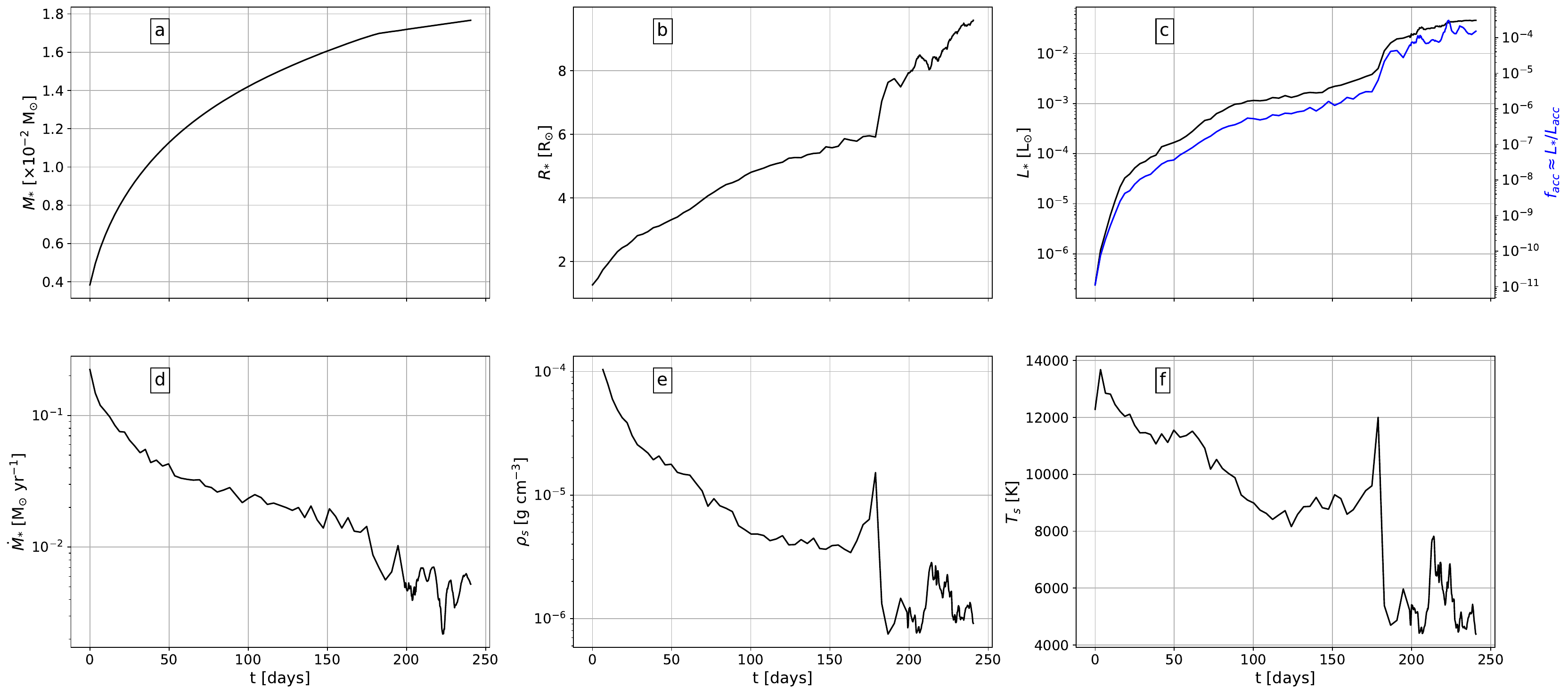}
    \caption{Evolution of the physical properties of the protostar displayed as a function of time, where $t=0$ marks the birth of the protostar. Panel (a) displays the protostar's mass, panel (b) its radius, panel (c) (resp. panel d) its surface integrated luminosity (resp. mass accretion rate), panel (e) (resp. panel f) the average density (resp. temperature) at the shock front. The solid blue line in panel (c) represents the radiative efficiency of the protostar.}
    \label{fig:protostarEvol}
\end{figure*}
\begin{figure*}
    \centering
    \includegraphics[width=1.\textwidth]{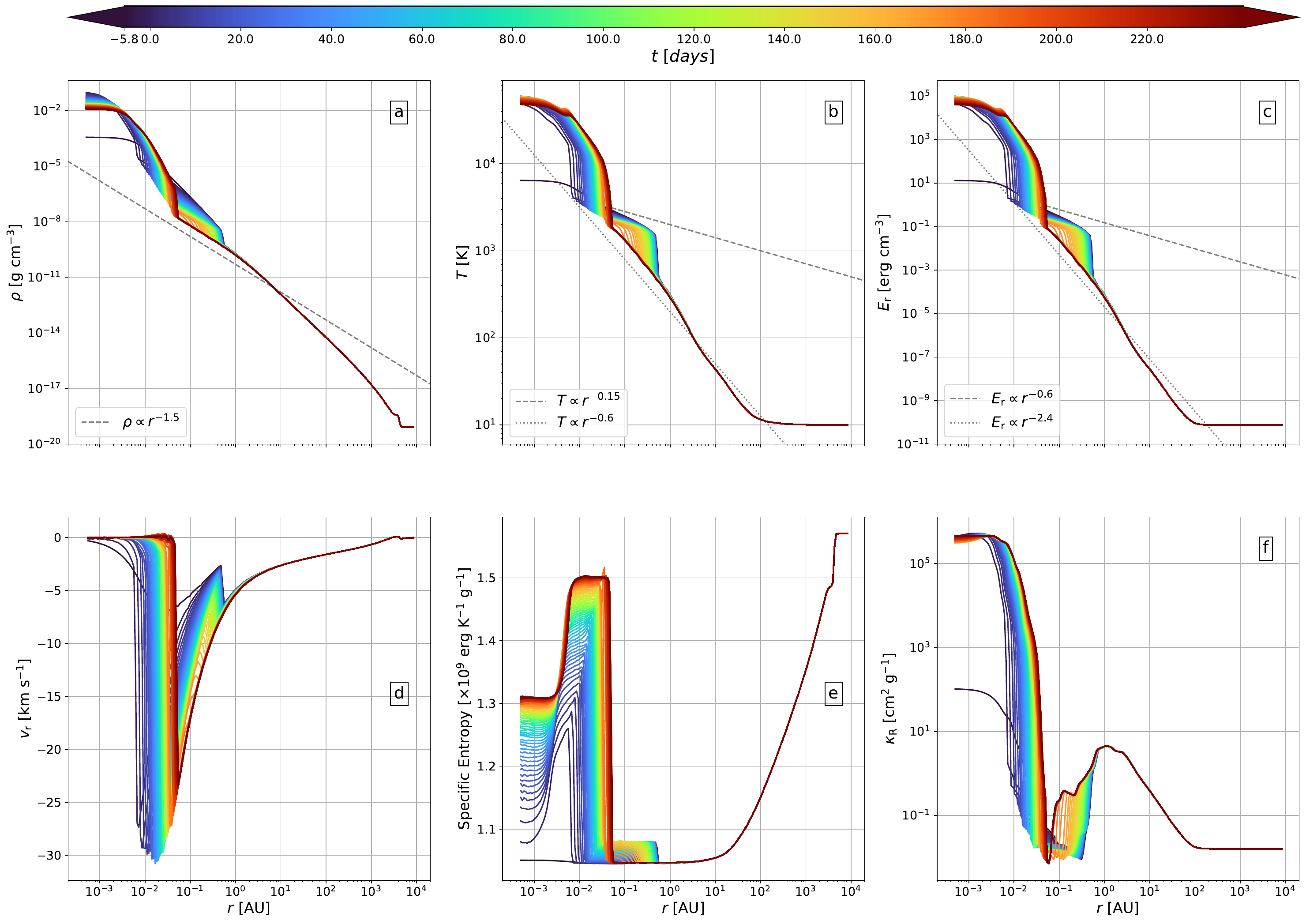}
    \caption{Evolution of the density (panel a), temperature (panel b), radiative energy (panel c), radial velocity (panel d), specific entropy (panel e), and Rosseland mean opacity (panel f) profiles, averaged in radial bins and displayed as a function of radius for different times, where $t=0$ marks the birth of the protostar. The last curves (dark red) on each panel correspond to $t\approx 241$ days. The dashed and dotted gray lines in panels (a), (b), and (c) are power law curves representing the expected density, temperature, and radiative energy profiles both prior to and after the accretion of the first Larson core.}
    \label{fig:G1Evol}
\end{figure*}

\begin{figure}
    \centering
    \includegraphics[width=.5\textwidth]{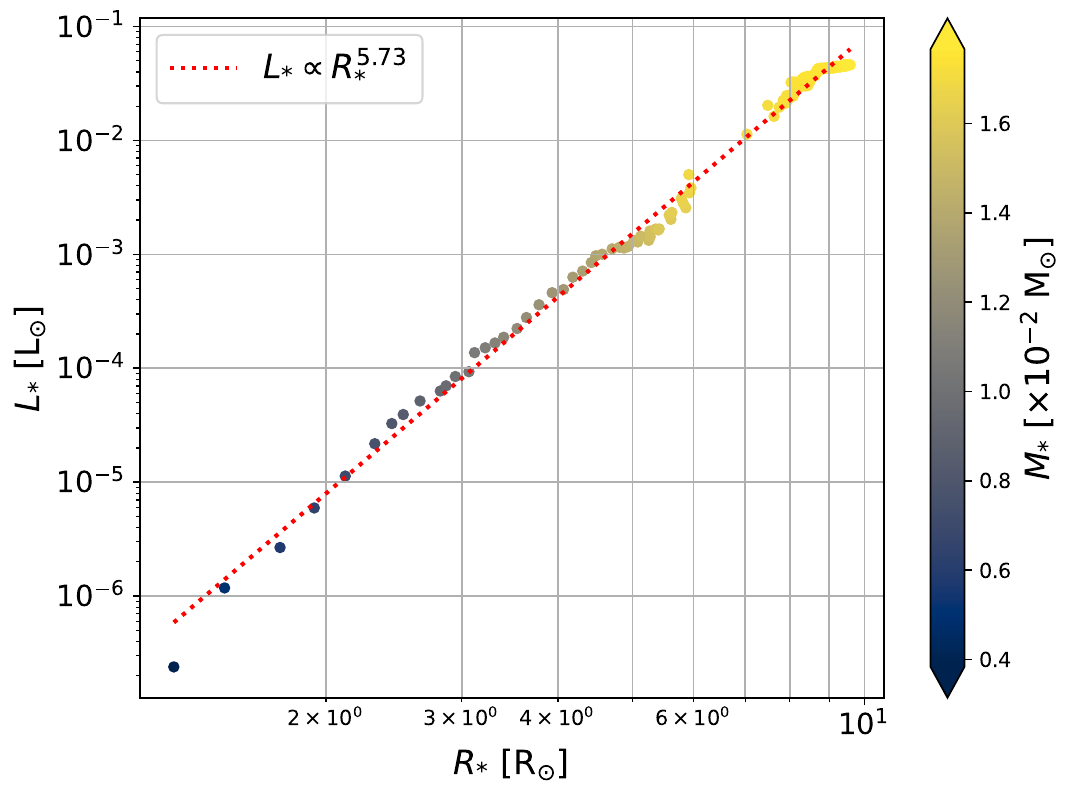}
    \caption{Logarithmic scatter plot showing the protostellar luminosity as a function of radius (where each scatter point is color coded with the protostellar mass). A fit (red dotted line) reveals a power law relationship between $L_{*}$ and $R_{*}$ whose exponent is $\approx 5.7$.}
    \label{fig:powerLaw}
\end{figure}

We now turn to studying the evolution of the protostar over time. Due to our high resolution, the time stepping is very stringent. In addition, we have $\sim 3\times 10^{7}$ cells inside the protostar's volume, which resulted in a very heavy computational load and our ability to integrate across long timescales was heavily impacted. Nevertheless, the results obtained provide us with valuable insights into the evolution of its physical properties and the radiative behavior of the accretion shock.
\\
\\
We thus begin by studying \hyperref[fig:protostarEvol]{Fig. \ref*{fig:protostarEvol}}, which displays the evolution of various properties of the protostar. In order to compute these physical properties, we selected all cells whose thermal pressure support outweighs incoming ram pressure (see \hyperref[appendix:defProtostar]{Appendix \ref*{appendix:defProtostar}}). In addition, we leverage the complementary information available in \hyperref[fig:G1Evol]{Fig. \ref*{fig:G1Evol}}, which displays various physical profiles, averaged in radial bins and displayed as a function of radius at different times.
\\
In \hyperref[fig:protostarEvol]{Fig. \ref*{fig:protostarEvol}}, panel (a) displays the enclosed mass inside the protostar. The protostar is born with a mass of $M_{*}\approx 4\times 10^{-3}$ $\rm{M_{\odot}}$, which steadily grows over time. The mass accretion rate, displayed in panel (d), is computed by integrating the mass flux on the protostar's surface:
\begin{equation}
    \dot{M}_{*} = -\int_{S_{*}} \rho v_{\mathrm{r}} dS\ ,
\end{equation}
where $S_{*}$ is the protostar's surface. The mass accretion rate begins at a tremendous $0.2$ $\rm{M_{\odot}\text{ }yr^{-1}}$, and quickly declines to $5.2\times 10^{-3}$ $\rm{M_{\odot}\text{ }yr^{-1}}$ by the last snapshot of our simulation. The radius of the protostar is displayed in panel (b). It is formed with a radius of $R_{*}\approx 1.3$ $\rm{R_{\odot}}$, and it continuously increases over time. In view of the fact that it contains such a small mass, the large radii seen in panel (b) are intriguing. Indeed, panels (a) and (b) show that the protostar contains $\approx 1.7\times 10^{-2}$ $\rm{M_{\odot}}$ in a radius of $9.5$ $\rm{R_{\odot}}$ by the end of the simulation. This initial bloating phase has previously been reported in the literature \citealp{larson1969, narita_1970, winkler_1980, bhandare_2020}, and is caused by the radiative behavior of the shock front. As can be seen in \hyperref[fig:EKL]{Fig. \ref*{fig:EKL}} panel (b), the accretion shock has a very high optical depth, and its radiation is immediately absorbed by the gas just upstream, which is also optically thick. As a result, the protostar faces immense difficulty radiating away the kinetic energy of the gas it accretes, the majority of which is dumped into the internal energy budget of the protostar. This is more readily seen in panel (c) of \hyperref[fig:protostarEvol]{Fig. \ref*{fig:protostarEvol}}, which displays the surface integrated luminosity $L_{*}$ (measured just upstream of the accretion shock) as well as the fraction $f_{\mathrm{acc}}$ of the accretion luminosity $L_{\mathrm{acc}}$ radiated away (blue curve of panel c). These two quantities are computed as
\begin{equation}
    L_{*} = \int_{S_{*}} \frac{c\lambda \nabla E_{\mathrm{r}}}{\rho \kappa_{\mathrm{R}}}dS\ ,
\end{equation}
\begin{equation}
    f_{\mathrm{acc}} \approx \frac{L_{*}}{L_{\mathrm{acc}}}\ ,
    \label{eq:facc}
\end{equation}
where
\begin{equation}
    L_{\mathrm{acc}} = \frac{GM_{*}\dot{M}_{*}}{R_{*}}\ .
\end{equation}
\hyperref[eq:facc]{Equation \ref*{eq:facc}} is only an approximation of the radiative efficiency of the shock front because $L_{*}$ also contains the cooling flux emanating from the protostar's interior, although we expect the latter to be very small due to the optical depths such radiation has to travel through. All throughout the simulation, the protostar is extremely dim and it radiates only a minute fraction of the accretion luminosity. The continuous increase in protostellar luminosity is due to two reasons; the expanding radiative surface, and the decrease in shock density (see \hyperref[fig:protostarEvol]{Fig. \ref*{fig:protostarEvol}} panel e), which reduces the optical depth of the accretion shock and facilitates the escape of radiation. Although the surface temperature of the protostar also decreases, its rate of decrease is not enough to reduce its luminosity output over time.
\\
This accumulation of energy can also be seen in \hyperref[fig:G1Evol]{Fig. \ref*{fig:G1Evol}}, which displays the evolution of various radial profiles over time. In panel (e) of this figure, one can see that the specific entropy of the gas downstream of the shock front is continuously increasing over time: the entire profile shifts upward as accretion progresses. However as the mass accretion rate decreases, the rate of increase in specific entropy also decreases. One can also see an increase in entropy in between the first and second core borders, caused by the radiation produced at the protostar's shock front.
\\
Another insight provided by this plot is the fact that the entropy continuously rises with the radius inside the protostar at all times during the simulation, meaning that it remains radiatively stable. Despite this, one can see a plateau develop just downstream of the second core shock front which is induced by the transport of heat in these regions. The mechanism behind this heat transport is the turbulent motion found within the protostar, which allows for a redistribution of energy throughout the protostar, and thus causes the entire entropy profile to shift upward. This becomes prominent over time as the effects of this turbulence begin to materialize (see \hyperref[sec:turb]{Sec. \ref*{sec:turb}} and the turbulence crossing time in \hyperref[fig:turbEfficiency]{Fig. \ref*{fig:turbEfficiency}}, panel c). As a consequence, the turbulent transport of energy becomes increasingly prominent over time in the protostar's outer layers. Having carried out our simulation in 3D, our more complete description of turbulence has allowed this plateau to develop on much smaller timescales than in \cite{bhandare_2020}'s 2D simulations (see \hyperref[appendix:bhandareComparison]{Appendix \ref*{appendix:bhandareComparison}}). One can also distinguish a second plateau develop in the innermost regions. This secondary plateau is caused by the high degree of ionization in the central regions (see \hyperref[fig:chemistry]{Fig. \ref*{fig:chemistry}}), which causes the fluid to transition to a lower entropy regime. The turbulent transport of heat within the protostar then causes this secondary plateau to develop. The mixing of entropy plays a crucial role in regulating the protostar's radius. Indeed, as radiative cooling struggles to evacuate the immense amount of energy being accreted by the protostar, turbulence aids this process by redistributing heat in its outer regions, thus alleviating the bloating.
\\
\\
Curiously, panel (b) of \hyperref[fig:protostarEvol]{Fig. \ref*{fig:protostarEvol}} displays a sudden increase in protostellar radius at $t\approx 187$ days, which coincides with a sudden increase and subsequent drop in shock density and temperature. This corresponds to a free fall time of the first Larson core, and indeed the various radial profiles in \hyperref[fig:G1Evol]{Fig. \ref*{fig:G1Evol}} confirm that the first core is accreted by the protostar at this moment (see for instance the disappearance of the first core accretion shock in panel a or d). Although this causes an order of magnitude increase in protostellar luminosity, the radiative efficiency remains well below unity since the protostar is still deeply embedded in an optically thick cloud. Its radius must further increase to larger values before the accretion shock can properly evacuate its radiative energy into a less dense and optically thinner medium.
\\
\\
\hyperref[fig:G1Evol]{Figure \ref*{fig:G1Evol}} also informs us of the behavior of the gas upstream of the protostar's shock front both prior to and after the accretion of the first Larson core. As mentioned previously, the density profile in the inner regions of the first Larson core follows $\rho\propto r^{-1.5}$ (gray dashed line) and $\rho\propto r^{-2}$ in its outer layers. As seen in panel (a) of the figure, the boundary between these two profiles expands outward over time, such that the entire density structure inside the first Larson core shifts to $\rho\propto r^{-1.5}$. This behavior has previously been reported by \cite{larson_1972, shu_1977}. The temperature profile follows $T\propto \rho^{\gamma_{\mathrm{eff}}-1}$. Prior to the accretion of the first core, the $\mathrm{H}_{2}$ molecules are undergoing the dissociation process, which places $\gamma_{\mathrm{eff}}$ at $\approx 1.1$. As a result, the temperature profile follows $T\propto r^{-0.15}$ (gray dashed line in panel b). Once the first core is accreted, the protostar directly accretes hot (and hence excited) $\mathrm{H}_{2}$ molecules, whose $\gamma_{\mathrm{eff}}$ is $\approx 7/5$. As a result, the temperature profile now shifts to $T\propto r^{-0.6}$ (gray dotted line in panel b). We see the same behavior in the radiative energies; since $E_{\mathrm{r}}\propto T^{4}$, we have $E_{\mathrm{r}}\propto r^{-0.6}$ prior to the accretion of the first Larson core, and $E_{\mathrm{r}}\propto r^{-2.4}$ afterwards.
\\
\\
Despite the nonlinear nature of the problem, it is our hope that a sub-grid model could be developed to properly describe the radiative feedback of the protostar unto its surrounding environment. To this end, we have displayed in \hyperref[fig:powerLaw]{Fig. \ref*{fig:powerLaw}} the protostar's surface integrated luminosity, plotted against its radius. This has demonstrated a power-law relationship between the two, where $L_{*}\propto R_{*}^{5.7}$. The power-law fit was performed prior to the accretion of the first core (i.e., $R_{*}<6\ \mathrm{R_{\odot}}$), as later times exhibit differing gas behaviors upstream of the accretion shock (\hyperref[fig:G1Evol]{Fig. \ref*{fig:G1Evol}}), which in turn changes the exponent of the power-law. In addition, we do not have a sufficient number of data points to accurately describe $L_{*}(R_{*})$ after the accretion of the first core. Although this result's robustness needs further testing and investigation, it excitingly hints at the existence of an analytical model that can be found. Such a model will need to describe the temporal evolution of the gas behavior both upstream and downstream of the shock front, whereby one estimates the amount of radiative flux escaping the protostellar surface based on the local gas structure. We plan to further explore this power-law relationship between $L_{*}$ and $R_{*}$ in the future.

\subsection{Chemical composition}

\begin{figure}
    \centering
    \includegraphics[width=.45\textwidth]{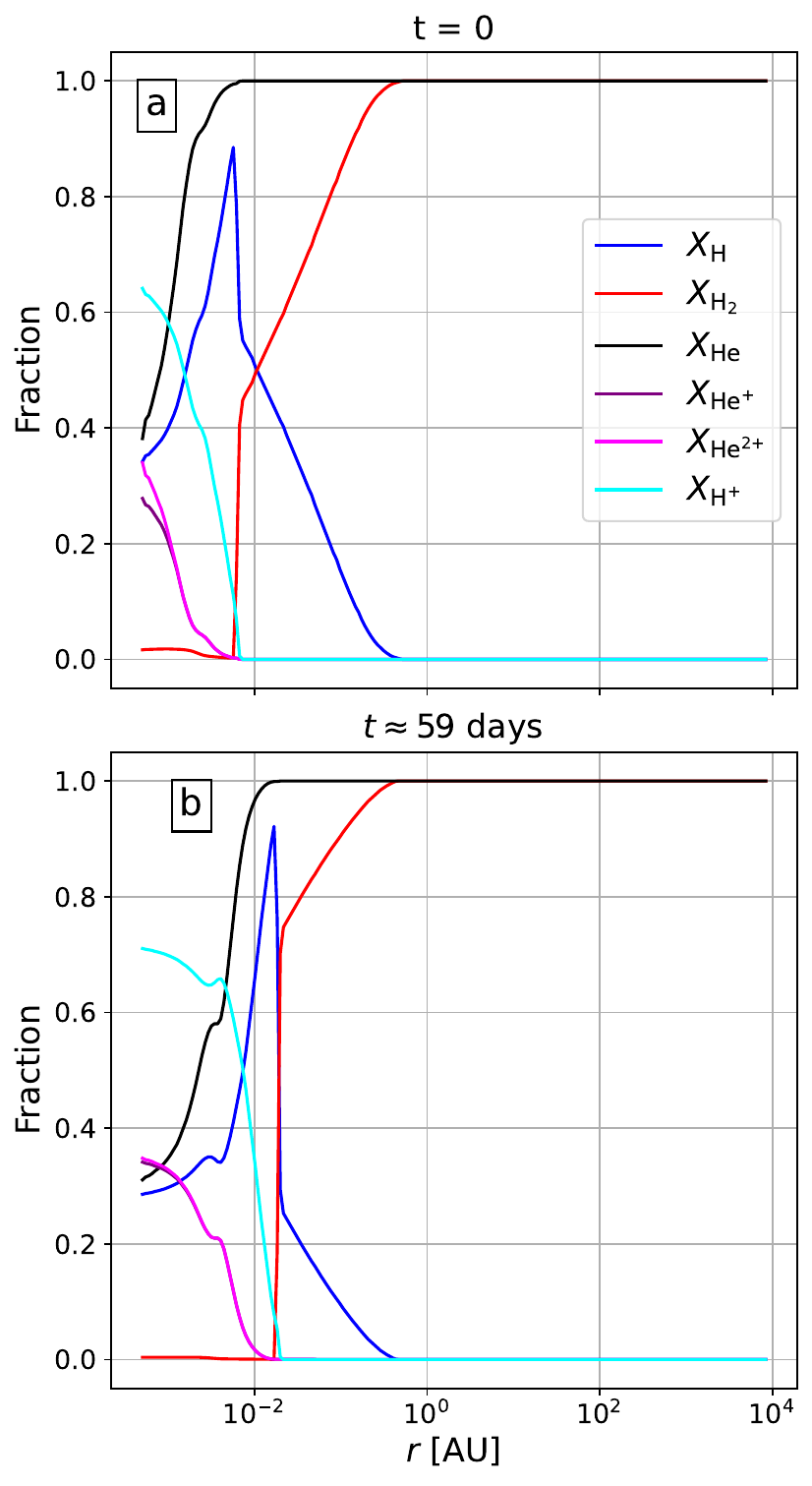}
    \caption{Fraction of $\mathrm{H}_{2}$ (red), $\mathrm{H}$ (blue), $\mathrm{H}^{+}$ (cyan), $\mathrm{He}$ (black), $\mathrm{He}^{+}$ (purple) and $\mathrm{He}^{2+}$ (pink), averaged in radial bins and displayed as a function of radius at the epoch of the protostar's formation (panel a) and $\approx$ two months later (panel b). See \hyperref[eq:chemfractions]{Eq. \ref*{eq:chemfractions}}.}
    \label{fig:chemistry}
\end{figure}

\begin{figure}
    \centering
    \includegraphics[width=.45\textwidth]{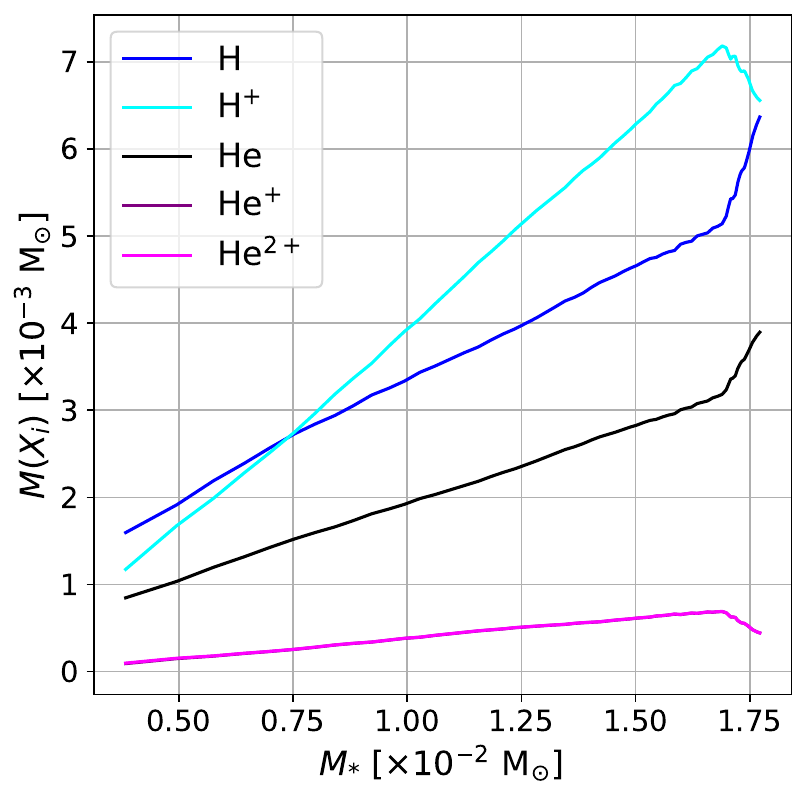}
    \caption{Mass of $\mathrm{H}$ (blue), $\mathrm{H}^{+}$ (cyan), $\mathrm{He}$ (black), $\mathrm{He}^{+}$ (purple) and $\mathrm{He}^{2+}$ (pink) inside the protostar displayed as a function of protostellar mass. The purple and pink curves overlap very closely.}
    \label{fig:chemistryEvol}
\end{figure}

An important factor to consider following our discussion in \hyperref[section:evol]{Sec. \ref*{section:evol}} is the dissociation of molecular hydrogen and the ionization of atomic hydrogen and helium. These processes consume energy, which is supplied by accretion and thus must be considered when attempting to determine the energy budget, and hence the radius of the protostar. The energy consumed by these processes is:
\begin{equation}
\label{eq:recipes}
\begin{aligned}
&\mathrm{H}_{2} \rightarrow 2\mathrm{H} : 4.48\text{ } \rm{eV}\ ,\\
&\mathrm{H} \rightarrow \mathrm{H}^{+} + \mathrm{e}^{-} : 13.60\text{ } \rm{eV}\ ,\\
&\mathrm{He} \rightarrow \mathrm{He}^{+} + \mathrm{e}^{-}: 24.59\text{ } \rm{eV}\ ,\\
&\mathrm{He}^{+} \rightarrow \mathrm{He}^{2+} + \mathrm{e}^{-}: 54.40\text{ } \rm{eV}\ .\\
\end{aligned}
\end{equation}
Using the equation of state table, we can directly estimate the fractions of each of these species in our computational domain by interpolating their values. As such, we do not actually model their dynamics, but simply provide the expected amount of each species for a given cell. We thus display in \hyperref[fig:chemistry]{Fig. \ref*{fig:chemistry}} the mass fraction of each species $\mathrm{X}_{i}$, averaged in radial bins where
\begin{equation}
\label{eq:chemfractions}
\begin{aligned}
&\mathrm{X_{H_{2}}}+\mathrm{X_{H}}+\mathrm{X_{H^{+}}} = 1\ ,\\
&\mathrm{X_{He}}+\mathrm{X_{He^{+}}}+\mathrm{X_{He^{2+}}} = 1\ .\\
\end{aligned}
\end{equation}
In panel (a) of this figure, we display these fractions at the epoch of the protostar's formation. A steep gradient in the fraction of $\mathrm{H}_{2}$ (red curves) is seen, which corresponds to the protostar's accretion shock. Here, all remaining $\mathrm{H}_{2}$ molecules are dissociated as a result of the shock heating, and only atomic hydrogen enters the protostar. The intense shock heating also begins ionizing the neutral hydrogen atoms (cyan curves), which happens in a much more gradual manner. However, the temperatures are not high enough to ionize the entirety of the atomic hydrogen reservoir, even in the central regions. We also see the onset of single (purple curves) and double (pink curves) $\mathrm{He}$ ionization just downstream of the accretion shock. The temperatures achieved in these regions cause a similar amount of $\mathrm{He}$ in first and second ionization states, although the curves begin to differ in the central regions. We see the same patterns $\approx$ 2 months later in panel (b), although the accretion shock has moved outward and the total fraction of ionized $\mathrm{H}$ has increased, whereas the fraction of ionized $\mathrm{He}$ remains the same.
\\
\\
Using these fractions, we also compute the mass of each of these species and display them in \hyperref[fig:chemistryEvol]{Fig. \ref*{fig:chemistryEvol}} as a function of the protostar's mass ($M_{*}$, analogous to time). Since almost no Hydrogen is under molecular form inside the protostar, we have omitted displaying the mass this species represents in the figure. We see an almost linear increase of all species with $M_{*}$, although the slopes for each species differs. At about $M_{*} \approx 7.5\times 10^{-3}$ $\rm{M_{\odot}}$, ionized Hydrogen becomes the dominant species inside the protostar in terms of mass, and by $M_{*} \approx 1.7\times 10^{-2}$ $\rm{M_{\odot}}$, about $\approx 50\%$ of the protostar's mass is under ionized form. However, the estimated amount of ionized material begins to decrease shortly afterward due to the decreasing density and temperature in the central core (see panels (a) \& (b) of \hyperref[fig:G1Evol]{Fig. \ref*{fig:G1Evol}}). In any case, this figure shows us that the electrical conductivity of the protostar remains high following its birth.
\\
By computing the total energy consumed by the dissociation and ionization processes, we find that they represent only $\approx 6\%$ of the total energy injected by accretion since the protostar's birth. As such, the rest of the accretion energy is either dumped into the internal energy budget of the protostar or used to drive turbulent motions, which are eventually converted into thermal energy. We estimate the fraction of the accretion energy used to drive turbulence in \hyperref[sec:turb]{Sec. \ref*{sec:turb}}.

\section{Turbulent motion within the protostar}
\label{sec:mainturb}
In this section, we aim to characterize the turbulence inside the protostar shown in \hyperref[fig:LIC]{Fig. \ref*{fig:LIC}} by describing it both quantitatively and qualitatively. We subsequently study how it evolves over time in our simulation.

\subsection{Onset of turbulence}
\label{sec:turbonset}
\begin{figure*}
    \centering
    \includegraphics[width=1\textwidth]{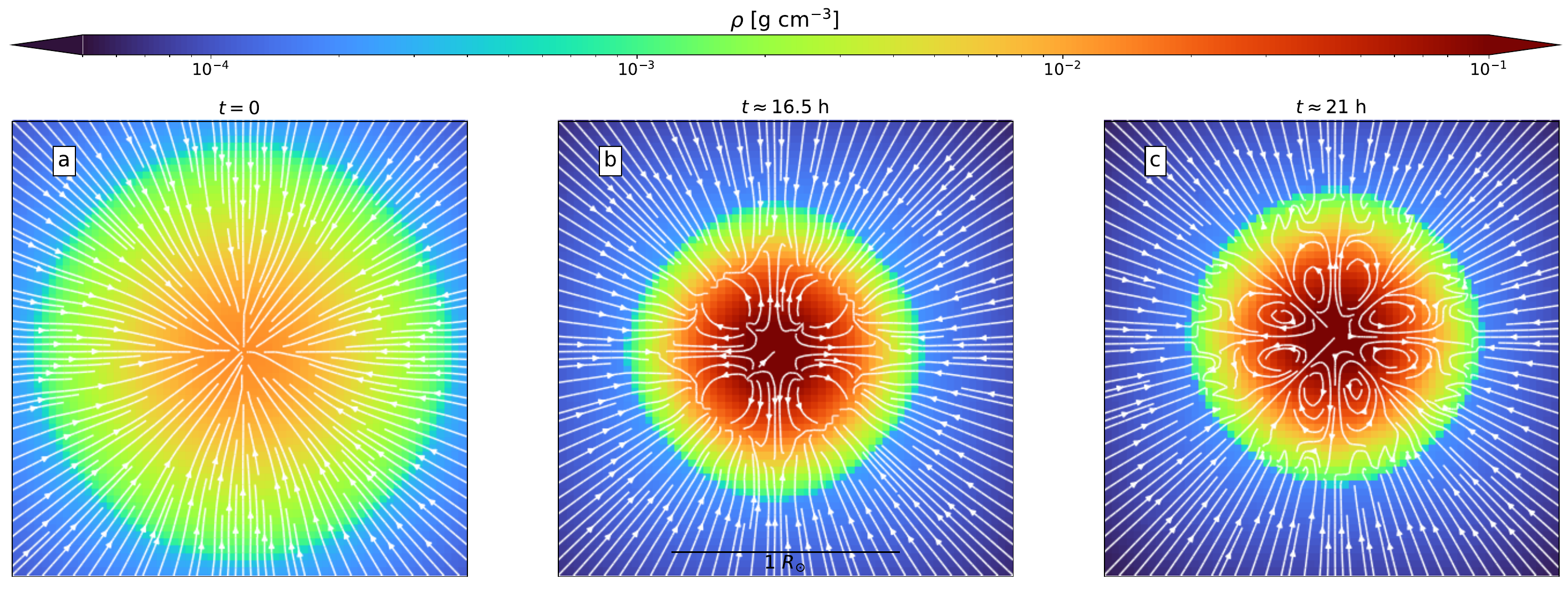}
    \caption{Density slices through the center of the domain showing the onset of turbulence within the protostar. Streamlines of the velocity vector field are shown in white. Each panel represents a different time, with panel (a) showing the protostar during the formation of the accretion shock ($t=0$), panel (b) after the onset of a hydro-dynamical rebound from the central region ($t\approx16.5$ h), and panel (c) after the outgoing wave interacts with the shock front ($t\approx 21$ h). The scale bar in panel (b) applies to the other two panels.}
    \label{fig:turbmechanism}
\end{figure*}

\begin{figure}
    \centering
    \includegraphics[width=.45\textwidth]{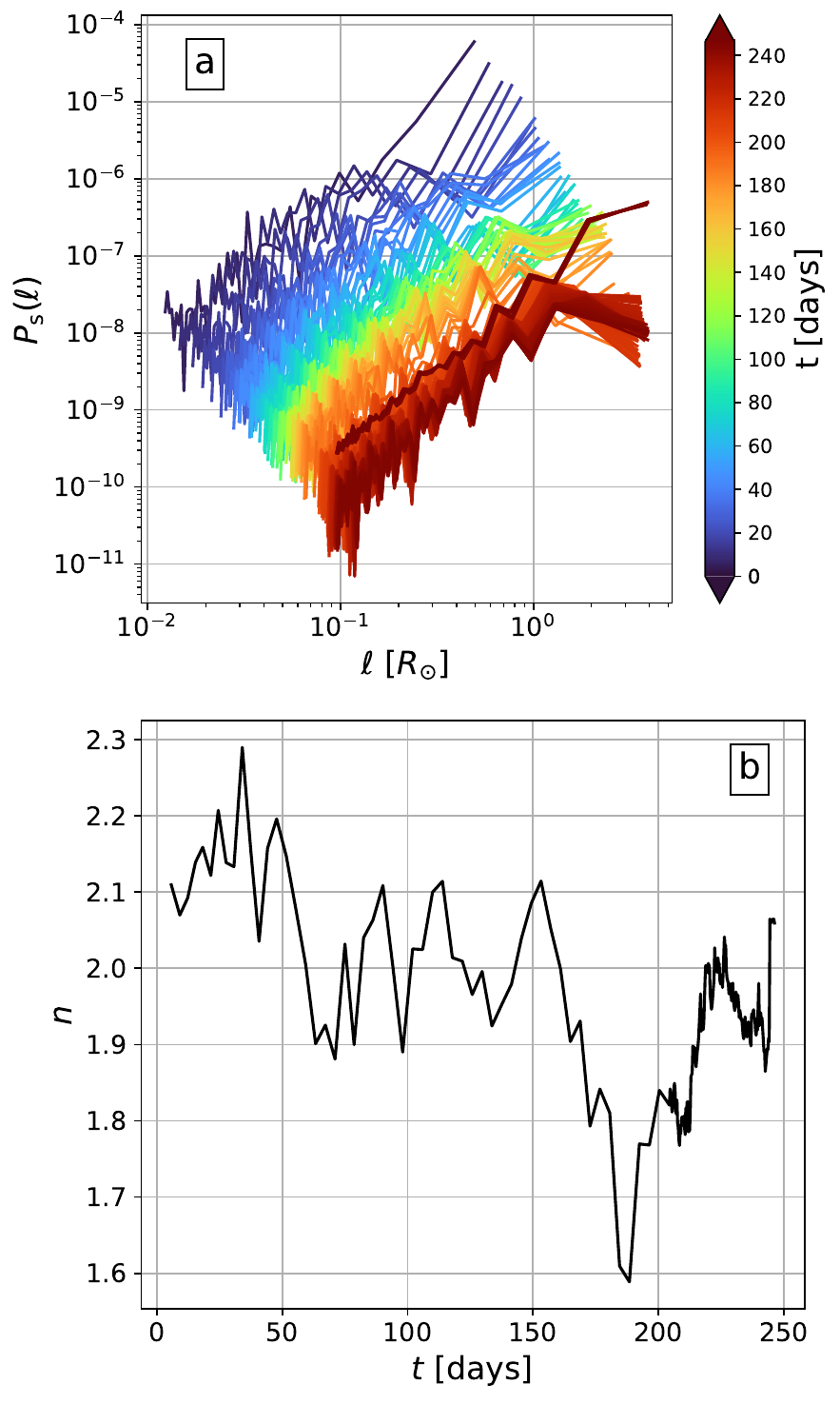}
    \caption{A spectral analysis of the turbulence within the protostar. Panel (a): Kinetic energy power spectrums as a function of characteristic scale $\ell$ of the gas within the protostar at different times, where $t=0$ marks the epoch of protostellar formation. The last curve (dark red) corresponds to $t\approx 241$ days. Panel (b): power-law fit of the curves in panel (a), where $P_{\mathrm{s}}(\ell)\propto \ell^{n}$.}
    \label{fig:powerspectrum}
\end{figure}

\begin{figure}
    \centering
    \includegraphics[width=.4\textwidth]{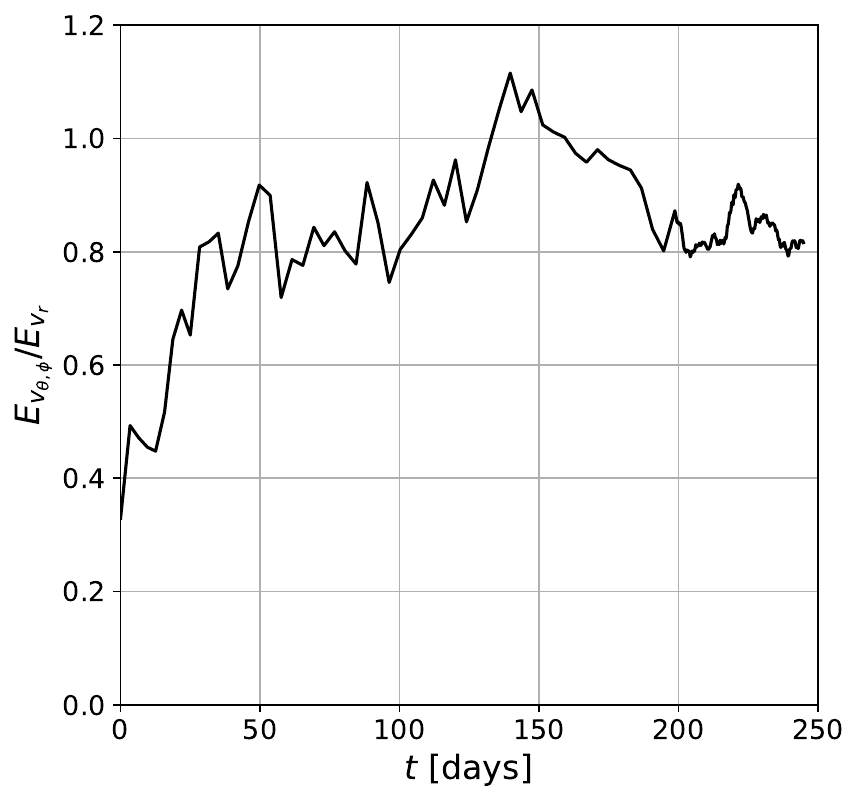}
    \caption{Ratio of ortho-radial to radial kinetic energy inside the protostar (see \hyperref[eq:kinorthorad]{Eq. \ref*{eq:kinorthorad}}) as a function of time, where $t=0$ marks the birth of the protostar.}
    \label{fig:turbgrowth}
\end{figure}

\begin{figure}
    \centering
    \includegraphics[width=.5\textwidth]{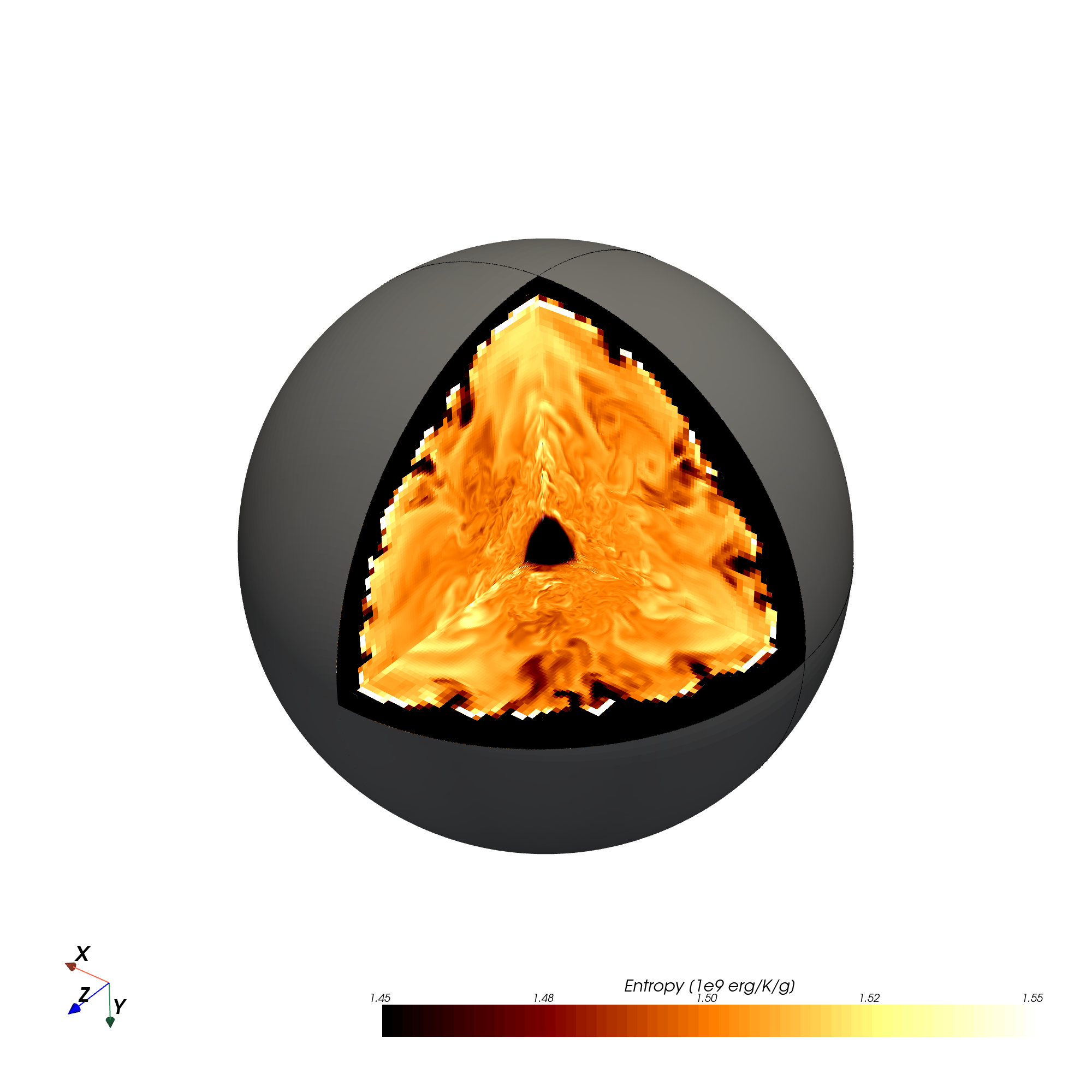}
    \caption{Cross-sectional view of the protostar at our final simulation snapshot, showing the interior entropy. The colorbar has been artificially anchored for visualization purposes. The gray spherical outline is an artistic choice for better visualization and serves no physical meaning.}
    \label{fig:fullstar}
\end{figure}

As stated previously, the rising entropy profile within the protostar suggests that this turbulence is not generated by a classical convective instability as postulated by Schwarzschild's criterion, where the protostar would exhibit a transition from a radiative zone to a convective shell. Thus, another instability seems to be at play here. Upon further investigation, we have discovered that the non-radial flow within the protostar has its origins during the hydrostatic bounce immediately following its formation.
\\
Indeed, \hyperref[fig:turbmechanism]{Fig. \ref*{fig:turbmechanism}} shows the protostar at different critical moments during its birth. Panel (a) shows the protostar as the second core accretion shock begins to form. Here, minute deviations from a purely radial flow can be seen downstream of the shock front. These are due to our use of a Cartesian grid, which favors flow along the grid axis. Upon crossing the shock front, the upstream velocity dispersions are amplified by about an order of magnitude, which allows them to be seen in the streamlines. Nevertheless, the kinetic energy carried by the non-radial flow is well below that of the radial flow.
\\
Several hours later, $\gamma_{\mathrm{eff}}$ reaches $4/3$ in the central regions owing to the rising density and temperature, thus forming a hydrostatic equilibrium that halts any further inward flow. This causes a hydrostatic bounce (panel b), where fluid with $v_{\mathrm{r}}>0$ \footnote{The outgoing wave is subsonic.} can be seen within the protostar. This bounce amplifies the non-radial flow within the protostar, although our grid geometry again seems to have an influence. Once the outgoing wave reaches the shock front, a physical instability seems to be triggered as strong vortical movement are produced within the protostar (panel c). Once the bounce has passed, these turbulent motions become sustained by accretion, as the supersonic radial flow of gas upstream of the accretion shock transfer's some of its momentum to the downstream gas, thus sustaining or amplifying any ortho-radial components in the downstream flow. This signals the onset of strong, stochastic turbulence within the protostar, as it becomes sustained through accretion. Indeed, \hyperref[fig:powerspectrum]{Fig. \ref*{fig:powerspectrum}} display the kinetic energy power spectrum $P_{\mathrm{s}}$ within the protostar throughout the simulation (panel a), which exhibits the power-law relationship governing $P_{\mathrm{s}}(\ell)$ and $\ell$, where $\ell$ is the inverse of the wavenumber. The exponent ($n$) of this power-law obtained through a numerical fit is displayed in panel (b); it hovers around 2. Although $n$ drops during the accretion of the first core when $t\approx 180$ days, it returns to 2 afterwards. This implies that the turbulence within the protostar is being continuously maintained by accretion.
\\
Another interesting observation provided by this figure is that the turbulence inside the protostar is not that expected of an incompressible fluid as postulated by \cite{kolmogorov_1941}, where $P_{\mathrm{s}}(\ell)\propto\ell^{11/3}$, despite the fact that the velocity dispersions are subsonic and well below the local sound-speed (blue curve in \hyperref[fig:turbCascade]{Fig. \ref*{fig:turbCascade}}). This is due to the heavily stratified nature of the protostellar interior, which hinders the inward motion of turbulent eddies.
\\
\\
In \hyperref[fig:turbgrowth]{Fig. \ref*{fig:turbgrowth}}, we compare the ortho-radial kinetic energy $E_{v_{\mathrm{\theta, \phi}}}$ with its radial counterpart $E_{v_{\mathrm{r}}}$ within the protostar. These two quantities are computed as:
\begin{equation}
\label{eq:kinorthorad}
    E_{v_{\mathrm{\theta, \phi}}} = 4\pi\int_{0}^{R_{*}} \rho r^{2} \left ( v_{\mathrm{\phi}}^{2} + v_{\mathrm{\theta}}^{2} \right ) dr, \\
    E_{v_{\mathrm{r}}} = 4\pi\int_{0}^{R_{*}} \rho r^{2} v_{\mathrm{r}}^{2} dr.
\end{equation}
Where $v_{\mathrm{\phi}}$ and $v_{\mathrm{\theta}}$ are respectively the azimuthal and meridional velocity. The curve suggests that the instability behind this turbulence causes an exponential growth of non-radial perturbations, before reaching a nonlinear phase where it stagnates. If equipartion is achieved, one would expect $E_{v_{\mathrm{\theta, \phi}}}/E_{v_{\mathrm{r}}} \approx 2$; however, the figure shows that the ratio reaches $\approx 0.8$ by $t\approx 30$ days and hovers around this value, meaning the flow within the protostar is mainly dominated by its radial component throughout the simulation.
\\
\\
In addition, although the entropy profiles averaged in radial bins in panel (e) of \hyperref[fig:protostarEvol]{Fig. \ref*{fig:protostarEvol}} show that the protostar is stable against convection, \hyperref[fig:fullstar]{Fig. \ref*{fig:fullstar}} shows that the turbulent motions can lead to local negative entropy gradients, where lower entropy fluid lies above higher entropy fluid. This causes weak convection to occur locally across all radii, and further contributes to the stochastic nature of the turbulence within the protostar.
\\
\\
We have seen this same pattern in higher resolution simulations; both $\ell_{\mathrm{max}} = 27$ and $\ell_{\mathrm{max}} = 28$ show the exact same onset of turbulence\footnote{We have presented the results of our $\ell_{\mathrm{max}} = 27$ in \hyperref[appendix:ResStudy]{Appendix \ref*{appendix:ResStudy}}; however, the time-stepping after second core formation in the $\ell_{\mathrm{max}} = 28$ was too stringent to produce any presentable results.}. Through private communications with A.Bhandare, we have learned that a similar phenomena seems to occur in \cite{bhandare_2020}'s 2D simulations run on a polar grid. Indeed, their protostar is turbulent at birth despite its radiative stability (see their Fig. C.1), and this turbulence begins following the hydrostatic bounce.
\\
When combining all of these elements together, we can conclude that although the seed for this turbulence has its origins in our grid geometry, the hydrostatic bounce and the subsequent amplification of turbulence caused by it and its interaction with the shock front are physical. We are still unsure as to what precise instability is at play here, but we have offered some evidence that could implicate the Standing Accretion Shock Instability (SASI, \citealp{blondin_2003, sheck_2004, foglizzo_2007}) in \hyperref[appendix:sasi]{Appendix \ref*{appendix:sasi}}.
\\
\\
In real astrophysical cases, the initial cloud core possesses both turbulent and rotational motion. If minuscule disturbances in the flow such as ours can provide the seed necessary to trigger turbulence within the protostar, then we predict that all protostars will be turbulent at birth.

\subsection{Accretion Driven Turbulence}
\label{sec:turb}

\begin{figure}
    \centering
    \includegraphics[width=.45\textwidth]{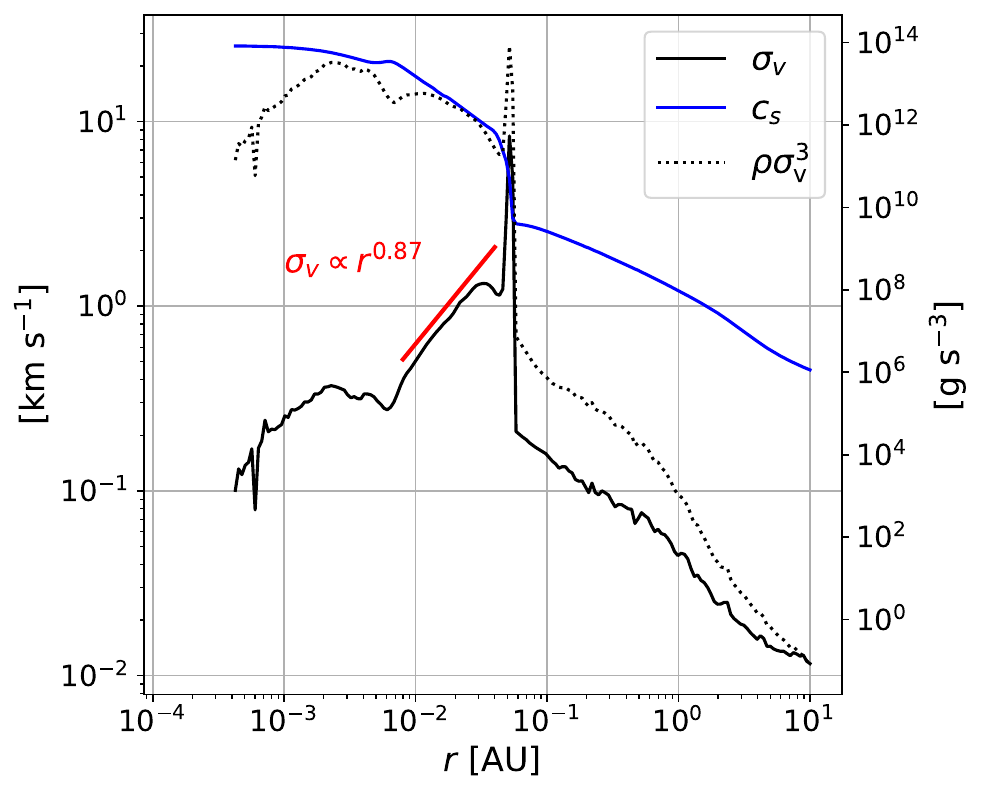}
    \caption{Velocity dispersion computed in radial bins (black curve) and average local sound speed (blue curve), displayed as a function of radius at our last simulation snapshot ($t\approx 241$ days, where $t=0$ marks the birth of the protostar). The red curve is a fit of the inertial range, whose exponent is $\approx 9/10$. The black dotted curve represents the turbulent energy flux (displayed in units of $\rm{g\text{ }s^{-3}}$).}
    \label{fig:turbCascade}
\end{figure}

\begin{figure}
    \centering
    \includegraphics[width=.5\textwidth]{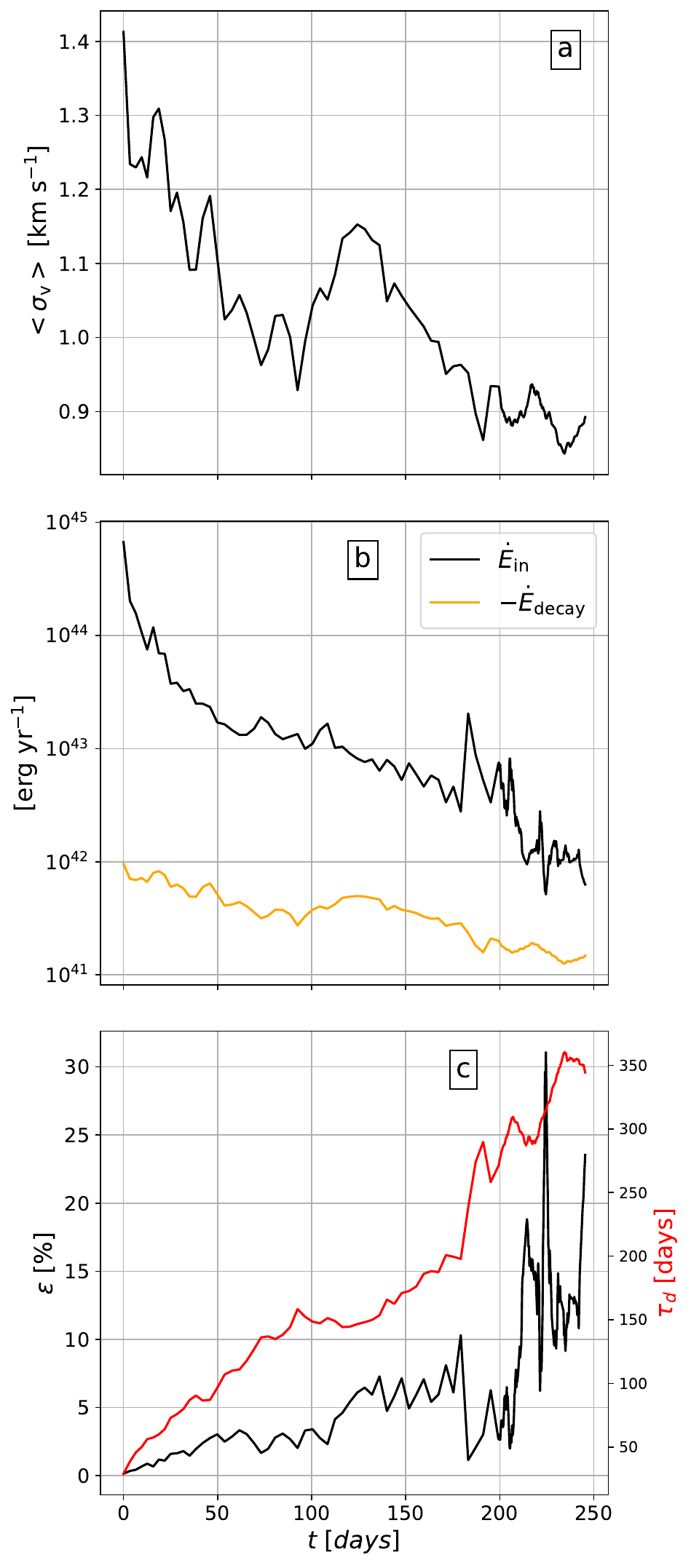}
    \caption{Mass-weighted velocity dispersion inside the protostar (panel a), injected accretion energy alongside the turbulence decay (panel b), and efficiency factor (panel c) displayed as a function of time, where $t=0$ marks the birth of the protostar. The red line in panel (c) corresponds to the turbulence crossing time (see \hyperref[eq:turbtimescale]{Eq. \ref*{eq:turbtimescale}}).}
    \label{fig:turbEfficiency}
\end{figure}

Now that we have established that turbulent motion is created at protostellar-birth and later sustained by accretion, we proceed by providing a quantitative analysis of its behavior throughout our simulation. To this end, we begin with \hyperref[fig:turbCascade]{Fig. \ref*{fig:turbCascade}}, which displays the velocity dispersions $\sigma_{\mathrm{v}}$ computed in radial bins as a function of radius (solid black line) at our last simulation snapshot. In this figure, the velocity dispersions upstream of the shock front are amplified by almost two orders of magnitude.
\\
Once the matter has properly settled into the protostellar surface, the velocity dispersions scale with the radius following a power-law $\sigma_{\mathrm{v}}\propto r^{9/10}$ (red fit in the figure). As the radius decreases, our ability to resolve these turbulent motions is hampered, since the number of cells in each radial bin decreases with decreasing volume. As a result, the scaling law is broken and the turbulence begins to dissipate through numerical diffusion. We would like to emphasize that the scaling law heavily depends on the internal structure of the protostar. As panel (a) of \hyperref[fig:G1Evol]{Fig. \ref*{fig:G1Evol}} has shown, the density profile (and hence the stratification) of the protostellar interior varies over time, which we have found is reflected in the scaling law between $\sigma_{\mathrm{v}}$ and $r$ (the proportionality exponent between $\sigma_{\mathrm{v}}$ and $r$ changes over time). Nevertheless, these turbulent motions carry a substantial amount of energy all throughout the protostar; the turbulent kinetic energy flux $\rho\sigma_{\mathrm{v}}^{3}$ (dotted line) remains strong all throughout the interior.
\\
\\
Since we are dealing with accretion driven turbulence, a fraction of the incoming accretion energy is used to drive turbulent motions inside the protostar. In order to determine this fraction, we base our analysis on the analytical tools provided by \cite{klessen_hennebelle}, which provides an estimate of the amount of turbulence generated by accretion and lost through decay in astrophysical bodies. Consequently, we begin by defining these tools, namely the turbulent crossing time $\tau_{\mathrm{d}}$, the turbulence driving scale which we assume to be $2R_{*}$, and the mean 3-dimensional velocity dispersion $\textlangle \sigma_{\mathrm{v}}\textrangle$ inside the protostar \citep{klessen_hennebelle}:
\begin{equation}
\label{eq:turbtimescale}
    \tau_{\mathrm{d}} \approx \frac{2R_{*}}{\textlangle \sigma_{\mathrm{v}}\textrangle}\ .
\end{equation}
One can also compute the amount of turbulent kinetic energy inside the protostar through
\begin{equation}
    E_{\mathrm{turb}} = \frac{1}{2}M_{*}\textlangle \sigma_{\mathrm{v}}\textrangle^{2}\ ,
\end{equation}
where $M_{*}$ is the protostar's mass. Using this, we can estimate the loss of turbulent kinetic energy over time $\dot{E}_{\mathrm{decay}}$
\begin{equation}
    \dot{E}_{\mathrm{decay}} \approx -\frac{E_{\mathrm{turb}}}{\tau_{\mathrm{d}}} = -\frac{1}{4}\frac{M_{*}\textlangle \sigma_{\mathrm{v}}\textrangle^{3}}{R_{*}}\ .
\end{equation}
Thus, in order to sustain the turbulence observed inside our protostar, it needs to be continuously driven by the incoming accretion energy $\dot{E}_{\mathrm{in}}$
\begin{equation}
    \dot{E}_{\mathrm{in}} = \frac{1}{2}\dot{M}_{*}v_{\mathrm{in}}^{2}\ ,
\end{equation}
where $v_{\mathrm{in}}$ is the infall velocity at the accretion shock. Finally, this allows us to compute the fraction of the accretion energy required to sustain the turbulence in the interior, which is characterized by the efficiency factor $\epsilon$:
\begin{equation}
    \epsilon = \left | \frac{\dot{E}_{\mathrm{decay}}}{\dot{E}_{\mathrm{in}}} \right |\ .
\end{equation}
If $\epsilon < 1$, then turbulence is sustained by accretion. In order to obtain $\textlangle \sigma_{\mathrm{v}}\textrangle$, we simply average the velocity dispersion inside the protostar by weighing it by mass. The mass weighing is done to ensure that the energy measurement is biased toward higher density gas.
\\
\\
In \hyperref[fig:turbEfficiency]{Fig. \ref*{fig:turbEfficiency}}, we display $\textlangle \sigma_{\mathrm{v}}\textrangle$, $\dot{E}_{\mathrm{in}}$, $\dot{E}_{\mathrm{decay}}$ and $\epsilon$ as a function of time. We have also displayed in panel (c) the turbulent crossing time (red line), which allows us to estimate the time required for the turbulence to dissipate from large eddies down to thermal energy. As the surface integrated mass accretion rate diminishes over time (see \hyperref[fig:protostarEvol]{Fig. \ref*{fig:protostarEvol}}, panel d), so too does the subsonic velocity dispersion inside the protostar. As a result, the accreted kinetic energy $\dot{E}_{\mathrm{in}}$ also reduces. The turbulence decay $\dot{E}_{\mathrm{decay}}$ also decreases over time. This is to be expected since the velocity dispersions decrease and the protostellar radius increases.
Regardless, the turbulence decay $\dot{E}_{\mathrm{decay}}$ remains well below the injected accretion energy at all times; the efficiency factor peaks at $\approx 31\%$. This shows that the injected accretion energy is abundant enough to sustain the observed turbulence inside the protostar at any point during the simulation. However, since the turbulent driving scale increases as the protostar grows, so too does the spatial extent of the turbulent cascade process. This is more readily seen in \hyperref[fig:LIC]{Fig. \ref*{fig:LIC}}, where larger eddies can be seen at the accretion shock as the protostar grows. This results in an increasing turbulent timescale, where the fraction of the injected accretion energy takes a more considerable amount of time to dissipate into thermal energy.
\\
\\
The ubiquitous turbulence found in the protostar raises the important question of how well it is described by our simulation. It is thus helpful to estimate the Reynolds number Re found within the protostar:
\begin{equation}
    \mathrm{Re} \sim \frac{2c_{\mathrm{s}}R_{*}}{v_{\mathrm{th}}\lambda_{\mathrm{p}}}\ ,
\end{equation}
where $c_{\mathrm{s}}$ is the sound speed, $\lambda_{\mathrm{p}}$ the particle mean free path, and $v_{\mathrm{th}}$ the thermal speed of hydrogen atoms:
\begin{equation}
    v_{\mathrm{th}} = \sqrt{\frac{3\kappa_{\mathrm{B}}T}{m_{\mathrm{H}}}}\ , \\
    \lambda_{\mathrm{p}} \sim 1/n\sigma\ ,
\end{equation}
where $n$ is the number density of atoms with collision cross-section $\sigma$ ($\approx 10^{-16}\ \mathrm{cm^{2}}$). By our simple estimates, the Reynolds number of the protostar's fluid should be $\sim 10^{14}$ at the surface ($c_{\mathrm{s}}\sim 1\ \mathrm{km\ s^{-1}}$, $T\sim 10^{3}\ \mathrm{K}$, $n\sim 10^{18}\ \mathrm{cm}^{-3}$) and $\sim 10^{17}$ in the central regions ($c_{\mathrm{s}}\sim 10\ \mathrm{km\ s^{-1}}$, $T\sim 10^{4}\ \mathrm{K}$, $n\sim 10^{22}\ \mathrm{cm}^{-3}$). These gargantuan Reynolds numbers mean that the characteristic scales by which viscosity effectively dissipates turbulence are orders of magnitude below our maximum spatial resolution. Indeed, such dissipation scales are on the order of the particle mean free path ($\sim [10^{-6}-10^{-3}$ $\rm{cm}]$), whereas our maximum spatial resolution is $\Delta x = 2.2\times 10^{9}$ $\rm{cm}$. As such, the turbulence is instead dissipated by our numerical diffusion, which means that our ability to describe this process is likely very impacted by our resolution. We have investigated the influence of our numerical resolution on the accretion driven turbulence in \hyperref[appendix:ResStudy]{Appendix \ref*{appendix:ResStudy}} and concluded that higher resolutions lead to stronger velocity dispersions in the protostar’s interior, which in turn amplifies the turbulent transport of heat. For further inquiries on turbulence in star formation related processes, we invite the reader to see \cite{Mckee_offner_2007, Hennebelle_2012}.

\section{Discussions} \label{section:discussions}

\subsection{The effects of initial conditions on the first and second Larson cores}

A result which initially intrigued us is the size of the first Larson core in our simulation. Indeed, \hyperref[fig:genesisHists]{Fig. \ref*{fig:genesisHists}} shows a first core radius of $0.5$ $\mathrm{AU}$. However, \hyperref[fig:EKL]{Fig. \ref*{fig:EKL}} shows a photosphere located at a much larger radius of $20$ $\mathrm{AU}$. As such, the location at which the fluid transitions from an optically thin regime to an optically thick one does not coincide with that of the first core border. This is despite the fact that isothermality is broken at the location of this transition (\hyperref[fig:genesisHists]{Fig. \ref*{fig:genesisHists}}, panel d). Hence, the radius of our first core is smaller than that which is commonly reported in the literature (e.g., \citealp{larson1969, vaytet_2013, vaytet_2017, bhandare_2018}). The small size of our first core can be attributed to our selection of the alpha value (\hyperref[eq:alpha]{Eq. \ref*{eq:alpha}}), which is smaller than those commonly adopted in the literature ($> 0.5$). For instance, \cite{vaytet_2013} compared the results of their simulations for different $\alpha$ values, and have found smaller first core radii for smaller $\alpha$ (see their tables 1 \& 2). This is due to the fact that smaller $\alpha$ values correspond to more violent gravitational collapses, where the high infall velocities and mass accretion rates lead to very strong ram pressure. As such, higher amounts of thermal pressure support are needed in order to attain a hydrostatic equilibrium in these configurations. 
\\
\\
The value of $\alpha$ that we have adopted has however little bearing on the subsequent formation of the protostar. Indeed, the high mass accretion rates unto the protostar (which begin at $\sim 10^{-1}\ M_{\odot}\ \mathrm{yr}^{-1}$ and decline to $\sim 10^{-3}\ M_{\odot}\ \mathrm{yr}^{-1}$ by our last snapshot) have previously been reported by numerous papers independently of the initial conditions and physical model adopted (e.g., \citealp{vaytet_2013, tomida_2013, Bate_2014, vaytet_2017, bhandare_2020}). The reason behind this is the first Larson core, which provides a momentary halt to accretion unto the central regions until temperatures can exceed $\approx 2000\ \mathrm{K}$, by which point the second collapse ensues. Since \cite{larson1969} has shown that the mass accretion rate asymptotically reaches $\sim c_{\mathrm{s}}^{3}/G$, then one can expect $\dot{M}_{*}\sim 10^{-2}\ M_{\odot}\ \mathrm{yr}^{-1}$, which explains the convergence seen in the literature.

\subsection{The radiative behavior of the protostar}
\hyperref[fig:protostarEvol]{Figure \ref*{fig:protostarEvol}} has shown us that the second core accretion shock remains subcritical throughout the simulation's duration, and as a consequence the protostar's radius swells dramatically over time. The most similar work in the literature to our study is that of \cite{bhandare_2020}, which also exhibits a substantial increase of the protostar's radius with mass. Since their study is two dimensional, they were able to integrate for much longer timescales (hundreds of years instead of our hundreds of days), and as such they were able to witness a contraction of the protostar in some of their simulations. This is explained by a reduction of the incoming mass accretion rate (i.e., a reduction in the incoming accretion energy), and an increased protostellar luminosity. They characterize this by comparing the Kelvin-Helmholtz timescale with the accretion timescale, which we have omitted from our study since the latter remains well below the former throughout our simulation\footnote{Our estimate of the radiative efficiency (\hyperref[eq:facc]{Eq. \ref{eq:facc}}) is equivalent to the ratio of the accretion timescale to the Kelvin-Helmholtz timescale.}. Once the Kelvin-Helmholtz timescale drops below the accretion timescale (i.e., $f_{\mathrm{acc}}>1$), the protostar can evacuate its energy, which causes the contraction. However, this occurs once the protostars have expanded to very large radii (on the order of a few AU, with a strong dependence on the initial cloud mass), where the accretion shock has reached first core densities. Furthermore, they do not evolve the simulations long after the contraction, meaning that it is unknown if this contraction is maintained all the way to the formation of a solar-like object. Nevertheless, the subcritical nature of the second core accretion shock has been widely reported in the literature \citep{larson1969, winkler_1980, vaytet_2013, vaytet_2018, Bate_2014, bhandare_2018, bhandare_2020}. It has been settled that the radiative efficiency of protostars must be high during most of its main accretion phase, as that would allow them to form with reasonably small radii \citep{larson_1972, appenzeller_1975, winkler_1980, stahler_1980}. Nevertheless, providing a quantitative estimate of the radiative efficiency of the second core accretion shock and how it varies over time remains of scientific interest. Indeed, many papers in the literature that are interested in larger spatial scales omit the expensive computations that we have performed; they set aside the protostar by replacing it with a sink particle, and prescribe its feedback effects using a sub-grid model (e.g.. \citealp{urban_2010, vorobyov_2015, hennebelle_2020, hennebelle_2022}). Thus, the radiative feedback of the protostar in these studies is $f_{\mathrm{acc}}\times L_{\mathrm{acc}}$, where $f_{\mathrm{acc}}$ is treated as a free parameter. The value of this parameter has been shown to have a significant effect on the resulting IMF \citep{hennebelle_2020}. Our simulation shows that the radiative efficiency is extremely low immediately following its birth, and although it increases significantly over time, it remains well below the current values used in the literature. However, we expect it to reach unity once most of the envelope has been accreted, as that would significantly reduce the optical depth of the shock front. This would subsequently allow the protostar to contract by radiating away the large amount of energy it has accumulated.

\subsection{The role of turbulence}

Regardless of our simulation's capacity in describing it, the existence of turbulent motion within the protostar from the moment of its inception is noteworthy, most notably for studies that aim to model the formation of stellar magnetic fields. Indeed, as stated previously, since dissipative effects such as ambipolar diffusion and ohmic dissipation considerably reduce the magnetic field strength implanted in the protostar, a dynamo process is required in order to generate the magnetic fields observed in young stellar objects ($\sim 1\ \mathrm{kG}$, \citealp{johns_2009}). In order to trigger such a dynamo process, convective motions are a prerequisite (e.g., \citealp{durney_1993, chabrier_2006}), and it is commonly believed that such motions arise once nuclear burning begins in the stellar core.  Indeed, the onset of nuclear fusion reverses the entropy profile inside the star, such that the central core will possess a higher entropy than the outer layers. This is due to the fact that the colossal amounts of energy generated by fusion can not be transported through radiation alone, and thus convective motions begin. Since our study has shown that turbulent motion emerges at protostellar birth, we reiterate \cite{bhandare_2020}'s hypothesis that a dynamo process can begin far earlier than previously thought. Since such a process draws from the kinetic energy budget of the protostar, then it can also participate in regulating its radius.

\subsection{Open questions}

In our opinion, our results raise important questions that we hope will be addressed in the future. Firstly, the manner in which the radiative behavior of the protostar differs when one includes more realistic initial conditions, where turbulence or solid body rotation in the initial dense molecular cloud core provide the angular momentum budget necessary to form a disk, should be thoroughly investigated. Although \cite{Bate_2014, vaytet_2018} have shown that the second core accretion shock remains strongly subcritical, \cite{vaytet_2018} has shown that the poles of the protostar radiate much more efficiently.
\\
Secondly, the extent with which turbulence helps in regulating the swelling of the protostar should be analysed in depth. Our resolution study has shown that higher resolutions lead to stronger velocity dispersions; however, since it is extremely difficult to further increase the resolution, we suggest that 1D calculations that include turbulence through mixing length theory might offer better insights in this regard (e.g., \citealp{larson1969, palla_1991}).
\\
Finally, magnetic fields can help in regulating the radius of the protostar, and a quantitative study in this regard is desirable. Indeed, previous studies in the literature have shown that magnetic fields can generate outflows (e.g., \citealp{machida_2006, machida_2007, tomida_2013, Bate_2014, tsukamoto_2015, wurster_2018, wurster_2020}, see also \citealp{mignon_2021} for the high mass case). Such outflows can extract a significant amount of energy which would have otherwise been accreted by the protostar.

\section{Conclusion}

We have carried out a simulation modeling the collapse of a gravitationally unstable, uniform density sphere of mass 1 $\rm{M_{\odot}}$ to protostellar densities, using a 3D RHD description of the gas dynamics under the FLD approximation. The calculations describe the initial isothermal phase, the first adiabatic contraction, the second gravitational collapse triggered by the dissociation of $\mathrm{H}_{2}$, and the second adiabatic contraction. We follow the evolution of the resulting protostar for $\approx 247$ days after its formation, which is longer than the first core free fall time of $\approx 187$ days and hence we were able to witness the latter's accretion by the protostar. Having placed a focus on the interior structure of the protostar, the simulation was carried out with the highest ever 3D resolution, which involved the use of 26 levels of refinement and $20-2\times 10^{3}$ cells per jeans length. Our findings can be summarized as follows:
\begin{enumerate}[label=(\roman*)]
  \item Following the formation of the protostar, its radius swells dramatically over time. This is due to the subcritical radiative nature of its shock front, which struggles to evacuate the immense amount of kinetic energy injected by accretion. The radiative efficiency of the protostar remains well below unity in the time-span that we have simulated, even after the accretion of the first core. However, as the protostar swells, the density (and hence the optical depth) of the accretion shock continuously decreases, which increases its radiative efficiency. We have revealed a power-law relationship between the luminosity just upstream of the shock front and the protostellar radius, a result which could aid in inferring the radiative behavior of the protostar across larger timescales once its robustness is established.
  \item Owing to our very high resolution, we were able to reproduce the findings of \cite{bhandare_2020}'s 2D simulations, where they have discovered that the protostar is turbulent from the moment of its inception despite its radiative stability. The turbulence is created during a hydrostatic bounce immediately following the birth of the protostar; it grows exponentially before reaching its nonlinear phase, where it is then maintained by accretion.. We have described this subsonic turbulence both quantitatively and qualitatively: a fraction ($< 31\%$) of the injected accretion energy is used to drive this turbulent motion, and the velocity dispersions show a power-law scaling with the radius. Since the protostar is heavily stratified, the behavior of this turbulence differs from the classical theory of \cite{kolmogorov_1941}. Due to the very high Reynolds numbers found in the protostar, our description of this turbulence is impacted by our numerical resolution. Our grid geometry also influences the behavior of the turbulence. Nevertheless, the heat transport it provides leads to significant entropy mixing and aids in regulating the protostellar swelling.
  \item We find that the protostar is not fully ionized at birth. However, as the protostar accretes material from its surroundings, the amount of mass within it under ionized form continuously increases over time. Hence, the electrical conductivity of the protostar increases over time. Additionally, we estimate that the dissociation of $\mathrm{H}_{2}$ and the ionization of atomic hydrogen and helium represents only $\approx 6\%$ of the total energy injected by accretion. As such, the energy consumption of these processes plays an insignificant role in regulating the radius of the protostar. Nevertheless, we predict that the high electrical conductivity of the protostar, when combined with the turbulence in the interior, could lead to a dynamo process prior to the onset of deuterium burning. Since generating the stellar magnetic field comes at the expense of kinetic energy, this could also aid in regulating the swelling of the protostar's radius.
  \item For the first time, we have carried out during these calculations a frequency-dependent treatment of radiative transfer. The results, presented in \hyperref[appendix:MultiGroup]{Appendix \ref*{appendix:MultiGroup}}, show no major differences to the gray approximation. This is in agreement with the 1D calculations of \cite{vaytet_2013}.
\end{enumerate}
Despite the short time-span of our simulation, we believe these results shed light on an otherwise poorly understood phase of the stellar formation process. We are currently investigating how this evolutionary picture changes once we include angular momentum in the system (which leads to the formation of a circumstellar disk), the results of which will be presented in a follow-up paper.

\begin{acknowledgements}
      We thank the anonymous referee for their useful comments that have improved the quality of this paper. This work has received funding from the French Agence Nationale de la Recherche (ANR) through the projects COSMHIC (ANR-20-CE31- 0009), DISKBUILD (ANR-20-CE49-0006), and PROMETHEE (ANR-22-CE31-0020). We have also received funding from the European Research Council synergy grant ECOGAL (Grant : 855130). We thank Thierry Foglizzo and Anaëlle Maury for insightful discussions during the writing of this paper. We also thank Asmita Bhandare for providing access to the data of their \cite{bhandare_2020} paper. The simulations were carried out on the Alfven super-computing cluster of the Commissariat à l'Énergie Atomique et aux énergies alternatives (CEA). Post-processing and data visualization was done using the open source \href{https://github.com/osyris-project/osyris}{Osyris} package.
\end{acknowledgements}

\bibliographystyle{aa}
\bibliography{biblio}

\begin{appendix}
\section{Defining the protostar in our simulation} \label{appendix:defProtostar}

\begin{figure}
    \centering
    \includegraphics[width=.45\textwidth]{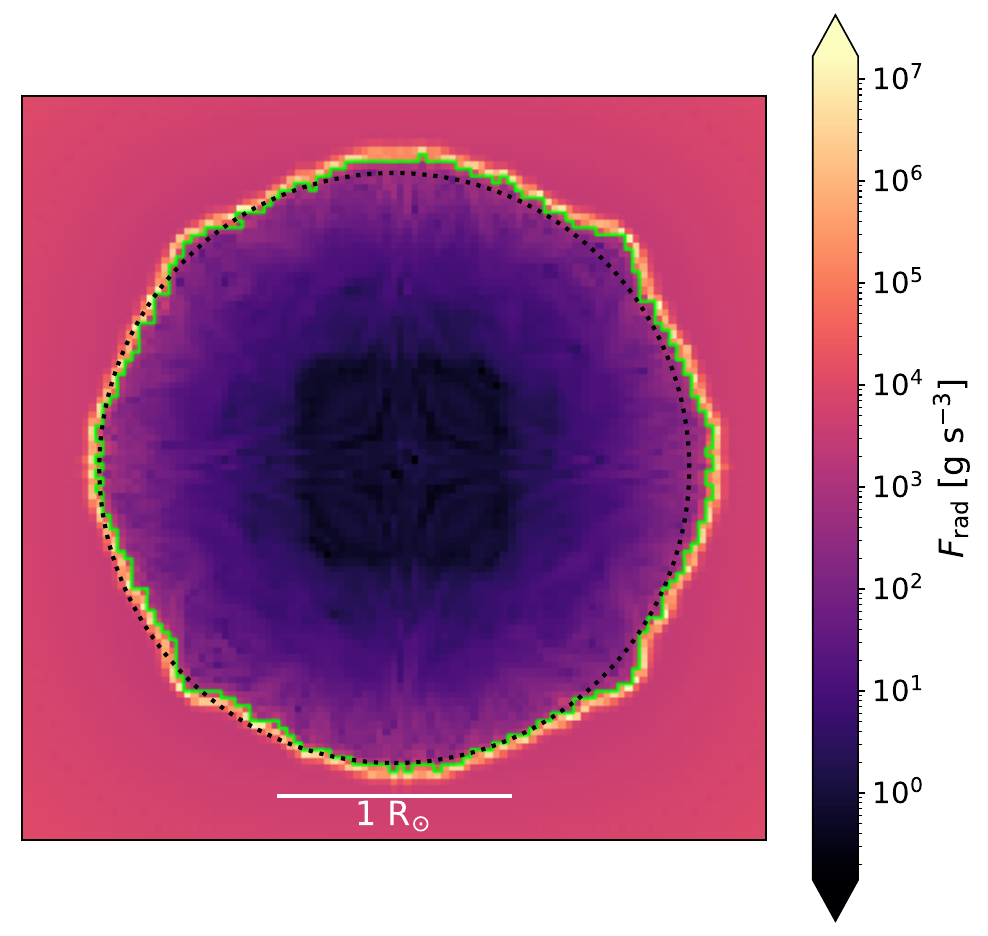}
    \caption{Illustration of our protostar definition criterion in a slice through the center of the domain at the epoch of protostellar formation. The colormap represents the local radiative flux, which prominently displays the second core accretion shock. The lime contour represents our $P\approx \rho v_{\mathrm{r}}^2$ criterion for the protostellar surface, of which all cells within it are counted as among the protostar. The dotted black circle represents an angular average radius of the protostellar surface.}
    \label{fig:definingProtostar}
\end{figure}

Herein, we present our definition of the protostar, namely the criterion by which we select cells that belong to it. Ideally, one would like to select all cells at and downstream of the accretion shock. For this, we have opted to adopt the criterion of \cite{tomida_2010}, which selects all cells whose thermal pressure support outweighs ram pressure ($P>\rho v_{\mathrm{r}}^2$). However, this criterion also selects cells belonging to the first core that are not currently undergoing a second gravitational collapse. As such, we have supplemented this criterion with a radius check, in which only cells at radii smaller than twice that of the $10^{-5}$ $\rm{g\text{ }cm^{-3}}$ density isocontour can be selected. In order to compute $R_{*}$, we simply average the radius of the $P\approx \rho v_{\mathrm{r}}^2$ contour (i.e., the protostellar surface). The results of this criterion are presented in \hyperref[fig:definingProtostar]{Fig. \ref*{fig:definingProtostar}}, which displays satisfactory results as the $P\approx \rho v_{\mathrm{r}}^2$ contour closely follows the accretion shock front.

\section{Resolution study} \label{appendix:ResStudy}

\begin{figure}
    \centering
    \includegraphics[width=.35\textwidth]{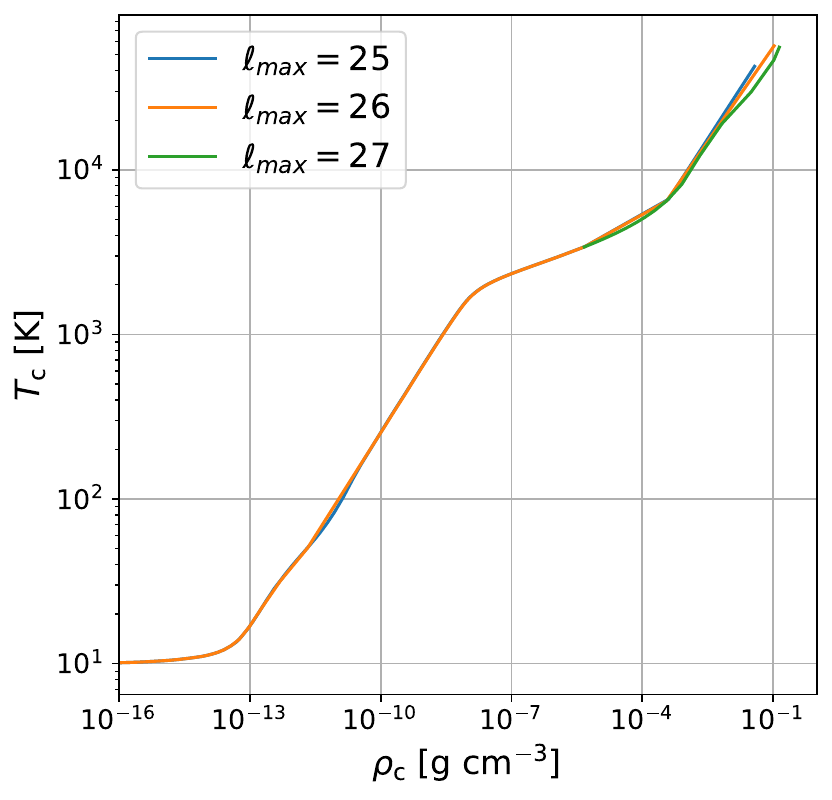}
    \caption{Central temperature plotted against central density prior to the first hydrodynamical bounce, for simulations with a maximum refinement level of 25 ($\Delta x = 2.93\times 10^{-4}$ $\rm{AU}$, blue curve), 26 ($\Delta x = 1.46\times 10^{-4}$ $\rm{AU}$, orange curve), and 27 ($\Delta x = 7.34\times 10^{-5}$ $\rm{AU}$, green curve).}
    \label{fig:rhocTc}
\end{figure}

\begin{figure}
    \centering
    \includegraphics[width=.35\textwidth]{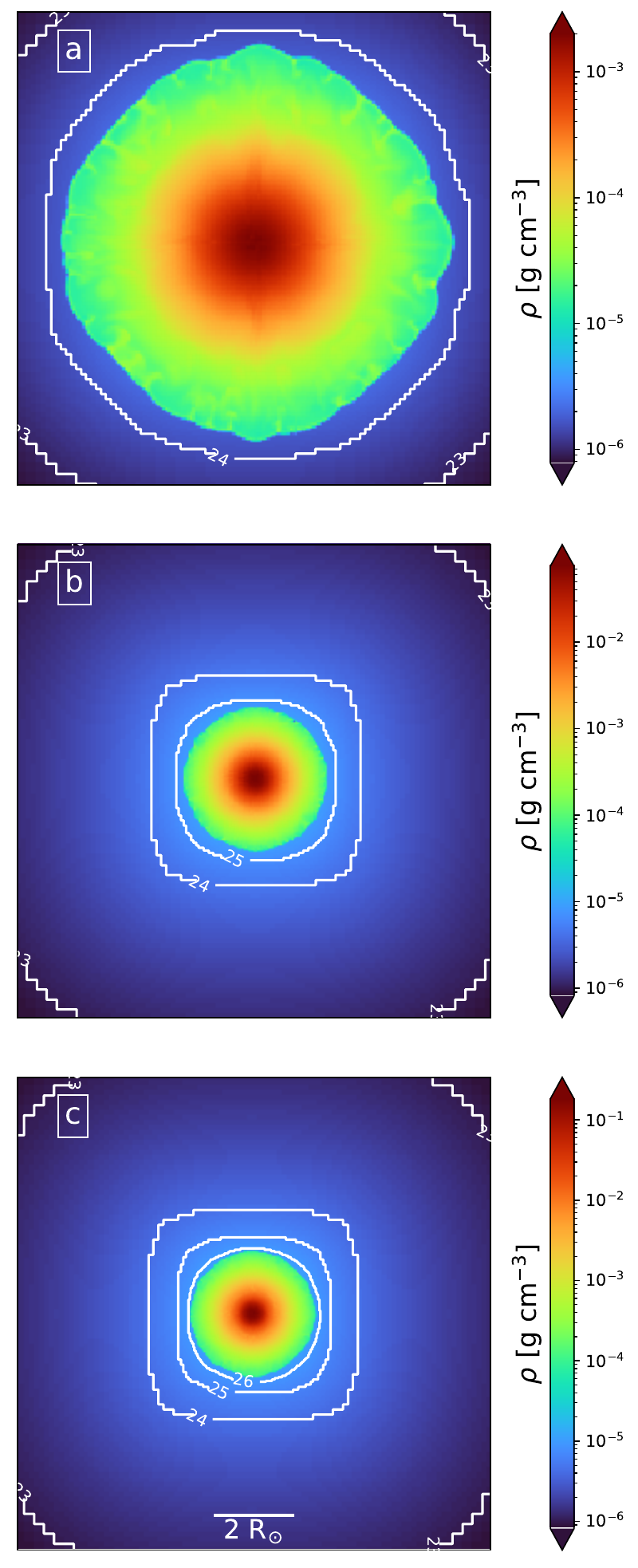}
    \caption{Density slices through the center of the domain for simulations with a maximum refinement level of 25 ($\Delta x = 2.93\times 10^{-4}$ $\rm{AU}$, panel a), 26 ($\Delta x = 1.46\times 10^{-4}$ $\rm{AU}$, panel b), and 27 ($\Delta x = 7.34\times 10^{-5}$ $\rm{AU}$, panel c). The scale bar in panel (c) applies to the other two panels. These slices are shown at a moment when all three protostars have reached similar masses ($5.9\times 10^{-3}$ $\rm{M_{\odot}}$ for panel (a), $5.75\times 10^{-3}$ $\rm{M_{\odot}}$ for panels (b) and (c)).}
    \label{fig:slicesResStudy}
\end{figure}

\begin{figure}
    \centering
    \includegraphics[width=.35\textwidth]{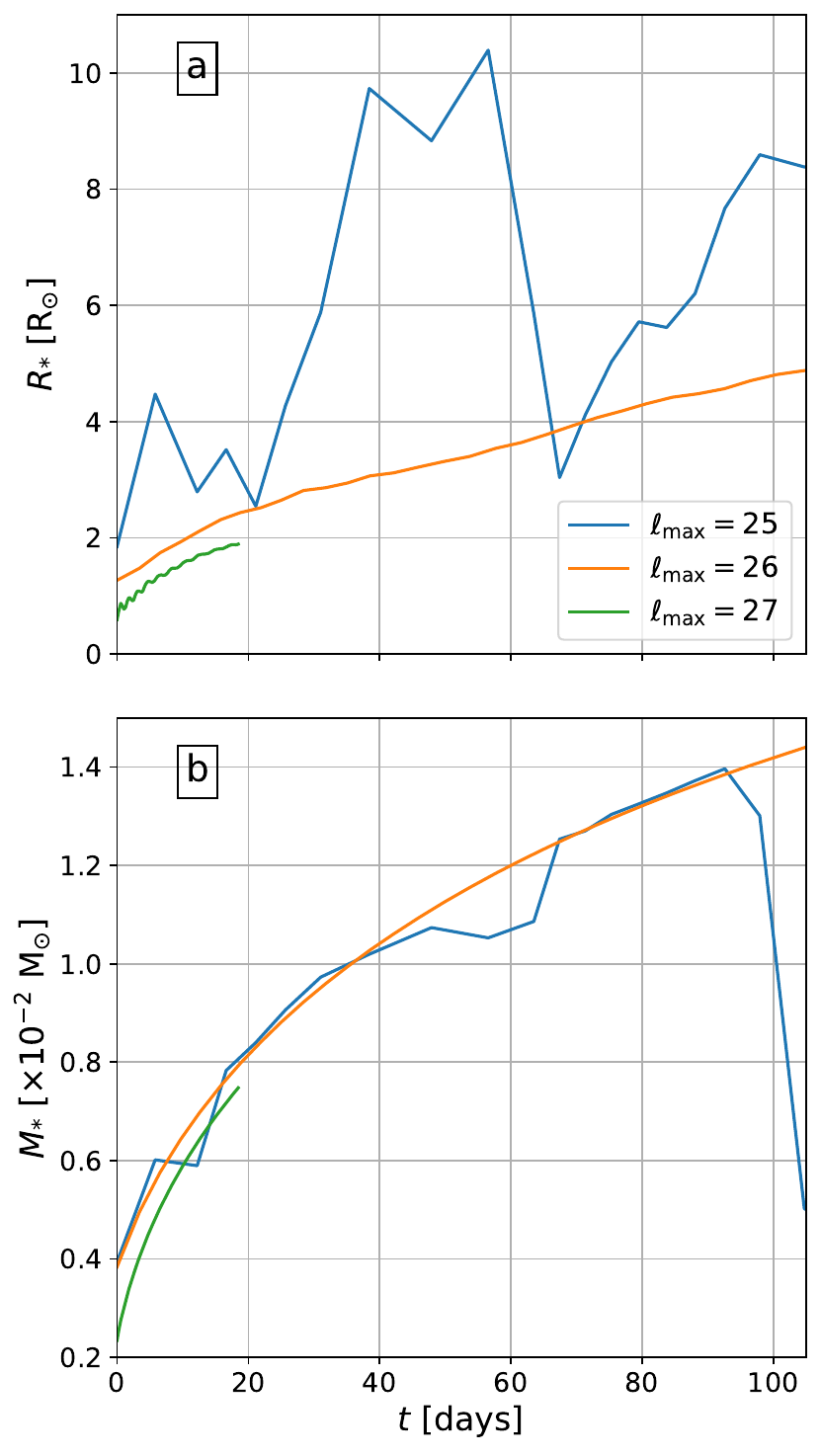}
    \caption{Radius (panel a) and mass (panel b) displayed as a function of time, where $t=0$ marks the epoch of protostellar birth, for simulations with a maximum refinement level of 25 ($\Delta x = 2.93\times 10^{-4}$ $\rm{AU}$, blue curve), 26 ($\Delta x = 1.46\times 10^{-4}$ $\rm{AU}$, orange curve), and 27 ($\Delta x = 7.34\times 10^{-5}$ $\rm{AU}$, green curve).}
    \label{fig:radiiMasses}
\end{figure}

\begin{figure}
    \centering
    \includegraphics[width=.35\textwidth]{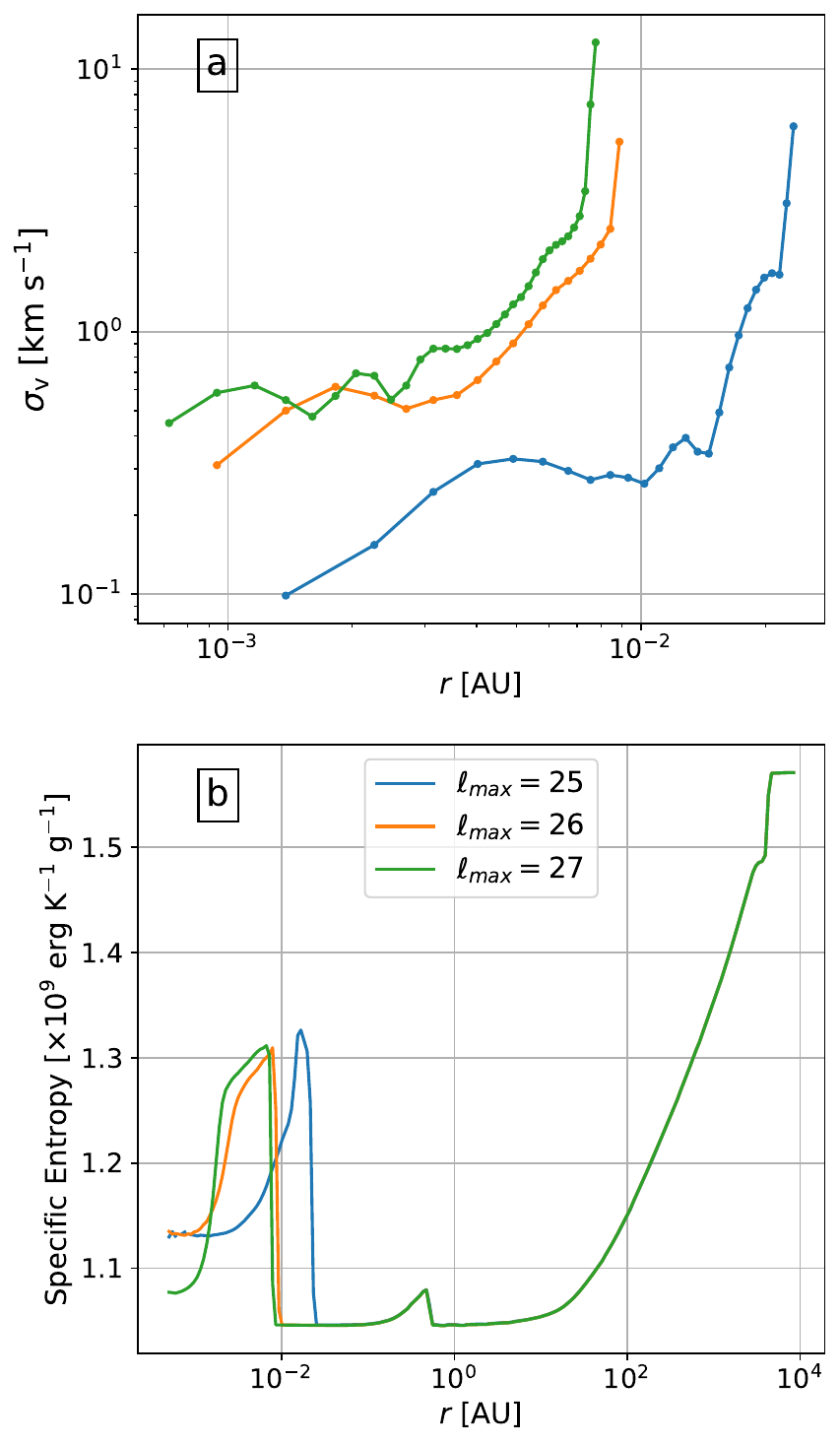}
    \caption{Velocity dispersion inside the protostar computed in radial bins (panel a) and average specific entropy (panel b), for simulations with a maximum refinement level of 25 ($\Delta x = 2.93\times 10^{-4}$ $\rm{AU}$, blue curves), 26 ($\Delta x = 1.46\times 10^{-4}$ $\rm{AU}$, orange curves), and 27 ($\Delta x = 7.34\times 10^{-5}$ $\rm{AU}$, green curves). These are shown at a moment in time where all three protostars have reached similar masses ($5.9\times 10^{-3}$ $\rm{M_{\odot}}$ for $\ell_{\mathrm{max}}=25$, $5.75\times 10^{-3}$ $\rm{M_{\odot}}$ for $\ell_{\mathrm{max}}=26$ and $\ell_{\mathrm{max}}=27$).}
    \label{fig:compareCascade}
\end{figure}

\begin{figure}
    \centering
    \includegraphics[width=.35\textwidth]{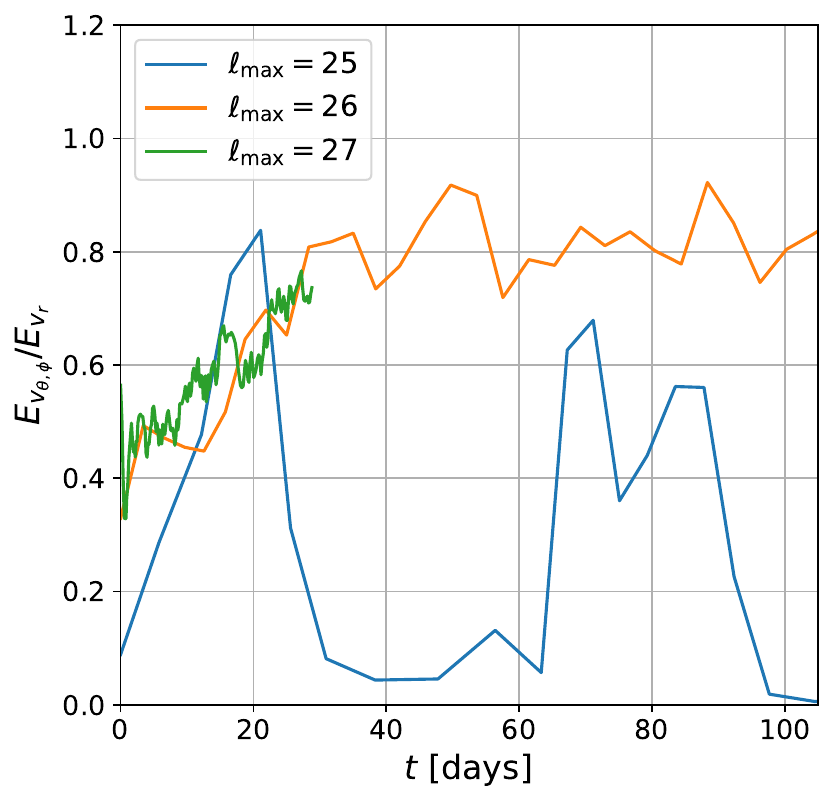}
    \caption{Kinetic energy of the ortho-radial flow compared to that of radial flow inside the protostar as a function of time, where $t=0$ marks the birth of the protostar, for simulations with a maximum refinement level of 25 ($\Delta x = 2.93\times 10^{-4}$ $\rm{AU}$, blue curve), 26 ($\Delta x = 1.46\times 10^{-4}$ $\rm{AU}$, orange curve), and 27 ($\Delta x = 7.34\times 10^{-5}$ $\rm{AU}$, green curve).}
    \label{fig:compareturbgrowth}
\end{figure}

As mentioned previously, simulating the stellar formation process requires a robust treatment of a multitude of physical processes. As such, each physical process requires an adequate spatial sampling in order to produce physical results. The most common approach in our field is a continuous refinement of the grid based on the local Jeans length as suggested by \cite{truelove_1997}. This study suggested that the Jeans length be resolved with at least four cells; however, this is inadequate to describe a second gravitational collapse, as the huge dynamical range requires a more diligent approach to spatial refinement. In addition, the inclusion of magnetic fields, be they ideal or nonideal, as well as the incorporation of radiative transfer can add further strain on simulations, as this requires additional spatial sampling to describe the full range of magnetic resistivities and the dust and gas opacities (see the discussions in \citealp{vaytet_2017, vaytet_2018, wurster_2022}). As such, it is important to carry out thorough examinations of the effect of resolution to test the convergence of each simulation based on the physical processes included in it, as well as the initial conditions with which it is carried out.
\\
\\
To this end, we have carried out two additional lower and higher resolution simulations in which we vary the maximum refinement level; however, the number of cells per $\lambda_{\mathrm{j}}^{*}$ was maintained at 20 (see Eq. \ref{eq:deltaX} and \ref{eq:jeansL}) as this has proven to be perfectly adequate. These two simulations possess a maximum refinement level $\ell_{\mathrm{max}}$ of 25 and 27, as opposed to the intermediate $\ell_{\mathrm{max}}=26$ of the simulation presented in the main body of this paper. This respectively offers them a spatial resolution of $\Delta x = 2.93\times 10^{-4}$ $\rm{AU}$ and $\Delta x = 7.34\times 10^{-5}$ $\rm{AU}$ at the maximal refinement level, as opposed to $\Delta x = 1.46\times 10^{-4}$ $\rm{AU}$.
\\
We thus show in \hyperref[fig:rhocTc]{Fig. \ref*{fig:rhocTc}} the central temperature as a function of the central density prior to the first hydrodynamical bounce (i.e., the moment when the central density drops from one snapshot to the next). The figure shows that prior to the formation of the protostar, all three simulations have followed identical evolutionary paths. However, the maximum density reached differs; the lower resolution run with $\ell_{\mathrm{max}}=25$ (blue curve) has attained $3.76\times 10^{-2}$ $\rm{g\text{ }cm^{-3}}$, the intermediate $\ell_{\mathrm{max}}=26$ (orange curve) reached $1.05\times 10^{-1}$ $\rm{g\text{ }cm^{-3}}$, and the higher resolution $\ell_{\mathrm{max}}=27$ (green curve) reached $1.39\times 10^{-1}$ $\rm{g\text{ }cm^{-3}}$. Thus, the central density achieved by the second gravitational collapse is resolution dependent; however, the intermediate resolution run achieved much closer results to the higher resolution run than to its lower resolution counterpart.
\\
We now turn to \hyperref[fig:slicesResStudy]{Fig. \ref*{fig:slicesResStudy}}, which shows density slices through the center of the domain for all three simulations with their AMR refinement level contours. These slices are shown at a moment in time where all three protostars have reached similar masses. Unsurprisingly, the spherical morphology of the protostar is better described in the intermediate (panel b) and high (panel c) resolution runs. Furthermore, the additional refinement levels allow a better resolution of the shock front, which is crucial to properly describe the sharp protostellar border. We also note that the lower resolution run displays a much larger radius than its intermediate and higher resolution counterparts.
\\
In \hyperref[fig:radiiMasses]{Fig. \ref*{fig:radiiMasses}}, we display the evolution of the radius (panel a) and masses (panel b) of the protostars. We note here that the higher resolution run (green curves) forms a smaller protostar, both in radius and in mass. In addition, it consistently shows smaller radii than the $\ell_{\mathrm{max}}=25$ and 26 runs at similar masses. The radius of the protostar in the lower resolution simulation fluctuates wildly, as the interior is poorly resolved in this run. In addition, the protostar in this run shows a huge, spurious drop in mass by $t\approx 90$ days, which demonstrates that it is inadequate to describe its evolution. Since the radiative behavior of the accretion shock front is identical in all three runs (i.e., extremely subcritical), the smaller radius in the higher resolution run is explained by the more adequate description of turbulence it provides. Indeed, we show in panel (a) of \hyperref[fig:compareCascade]{Fig. \ref*{fig:compareCascade}} the velocity dispersions computed in radial bins inside the protostar. The higher resolution run displays stronger velocity dispersions than the other two runs, which provides a better turbulent transport of heat. As a result, the plateau in the entropy profile is better developed here than in $\ell_{\mathrm{max}}=25$ and 26 runs, which shows that the energy has been better redistributed. Hence, the radius of the protostar in the higher resolution run is consistently smaller.
\\
Finally, we display in \hyperref[fig:compareturbgrowth]{Fig. \ref*{fig:compareturbgrowth}} the ratio of ortho-radial to radial kinetic energies (see \hyperref[eq:kinorthorad]{Eq. \ref*{eq:kinorthorad}}) of the protostars as a function of time. The temporal evolution here is similar for the $\ell_{\mathrm{max}}=26$ and 27 runs, but the $\ell_{\mathrm{max}}=25$ once again appears to be incapable of properly describing the turbulent motions within the protostar.
\\
\\
In summary, although this resolution study has shown that our simulations are not converged, the differences between the $\ell_{\mathrm{max}}=26$ and $\ell_{\mathrm{max}}=27$ runs are small enough for us to conclude that our results are sufficiently realistic for physical interpretations. When taking into account the stringent time-stepping involved in the $\ell_{\mathrm{max}}=27$ run (which we could not integrate past a dozen days), we have concluded that $\ell_{\mathrm{max}}=26$ is the optimal resolution choice.

\section{Collapse with multigroup radiative transfer} \label{appendix:MultiGroup}

\begin{figure*}
    \centering
    \includegraphics[width=0.9\textwidth]{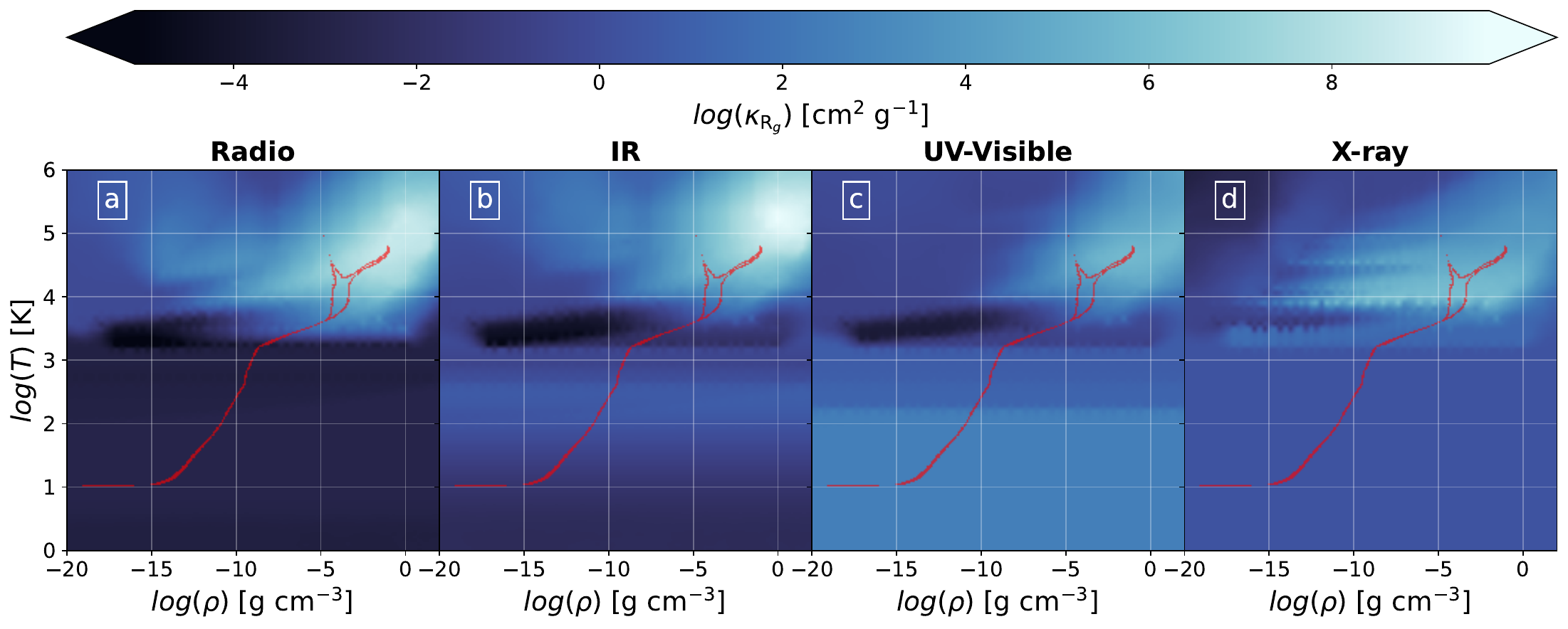}
    \caption{Rosseland mean opacity meshes created for each radiative group in our multigroup simulation (see \hyperref[tab:radGroups]{Table \ref*{tab:radGroups}}). The temperature-density distribution of all cells during the epoch of protostellar birth is overlaid in red.}
    \label{fig:M1Mesh}
\end{figure*}

\begin{figure}
    \centering
    \includegraphics[width=.45\textwidth]{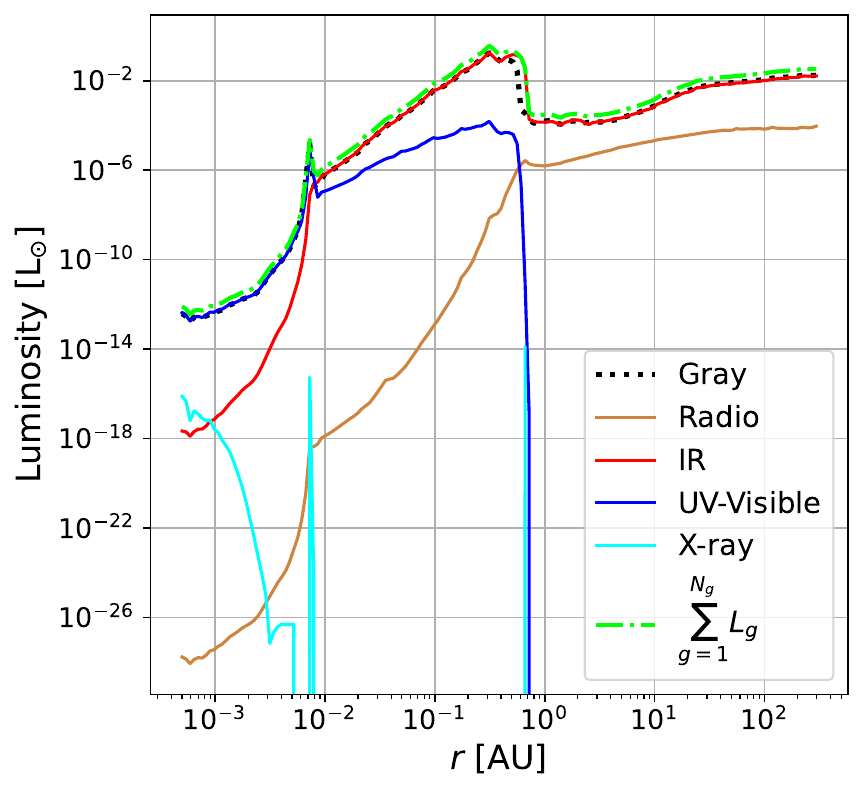}
    \caption{Luminosity profiles displayed as a function of radius at the epoch of the protostar's birth for the gray radiative transfer simulation (dotted black line) and its multigroup counterpart (colored solid lines). The lime dash-dotted line is the total luminosity in the multigroup run.}
    \label{fig:G1M1Lum}
\end{figure}

\begin{figure}
    \centering
    \includegraphics[width=.45\textwidth]{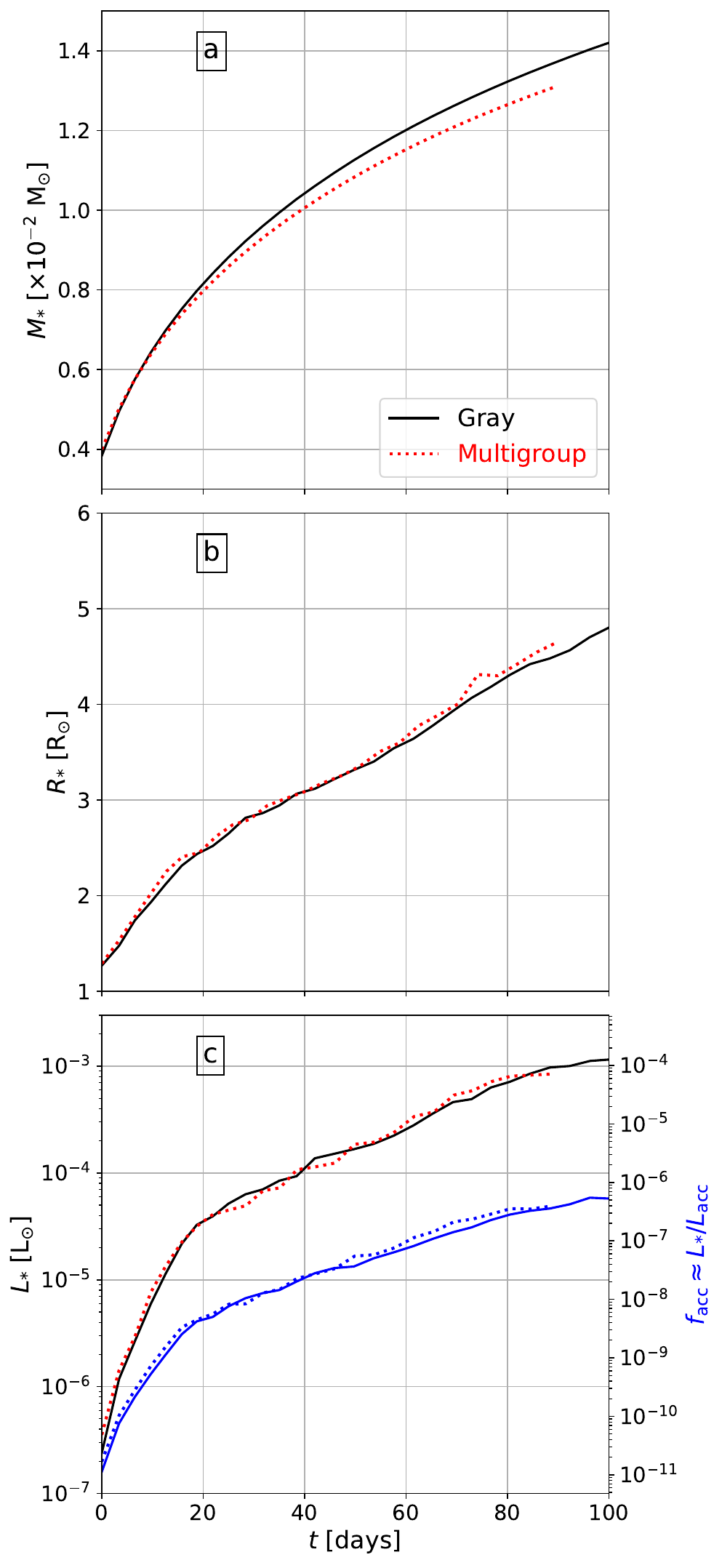}
    \caption{Comparison of the protostellar mass (panel a), radius (panel b), and luminosity (panel c) between our gray radiative transfer simulation (solid black lines) and its multigroup counterpart (dotted red lines). The solid (resp. dotted) blue line in panel (c) represents the protostellar radiative efficiency in the gray (resp. multigroup) simulation.}
    \label{fig:G1M1Evol}
\end{figure}

In order to test the robustness of our simulation's results, which uses a gray approximation for its radiative transfer, we have conducted a second simulation with a multigroup description. It possesses the same initial conditions; however, we now split the $[10^{5};10^{19}$ $\rm{Hz}]$ frequency range into four distinctive groups. The results of the simulation are extremely similar to that of its gray counterpart, as previously reported by the 1D calculations in \cite{vaytet_2013}.
\\
Our choice of 4 groups was the result of significant memory constraints. Indeed, since our maximum refinement level is 26, this requires the allocation of around 1.5 TB of RAM memory, of which $\approx$ 915 GB are used by the AMR grid. Furthermore, such a memory cost meant that the 64 processing cores had to be spread across 4 times as many computing nodes, which increased the CPU communications burden. In addition to the heightened computational load, the added communications burden constrained our ability to integrate the simulation on longer timescales.
\\
Since protostars form with temperatures $>10^{4}$ $\rm{K}$, most of the radiative energy is in the ultraviolet part of the electromagnetic spectrum. This energy is later absorbed by the surrounding gas and reemitted in the infrared. We thus chose to have both an infrared and an ultraviolet-visible group, with two other radiative groups bordering these two to avoid any energy omissions. The frequency borders of each group are presented in \hyperref[tab:radGroups]{Table \ref*{tab:radGroups}}, and the opacity meshes created for each of them using the previously mentioned Delaunay triangulation process are presented in \hyperref[fig:M1Mesh]{Fig. \ref*{fig:M1Mesh}}.
\\
\\
These meshes are very similar to the gray mesh (see \hyperref[fig:OPACITYMESH]{Fig. \ref*{fig:OPACITYMESH}}); the dust destruction front and the subsequent atomic opacity peak are both clearly visible except for the X-ray mesh (panel d), where the destruction of the dust particles barely has a noticeable effect. In addition, the X-ray mesh also contains a batch of triangles at log($T)\sim 3.5$, which is due to a lack of sampling points in the \cite{vaytet_2013} dataset. However, since this radiative group is the least prominent in terms of radiative energies, this will have a minor effect on our simulation.
\\
\\
The results of this multigroup simulation are displayed in Figures \hyperref[fig:G1M1Lum]{\ref*{fig:G1M1Lum}} and \hyperref[fig:G1M1Evol]{\ref*{fig:G1M1Evol}}, where we compare it with its gray counterpart. \hyperref[fig:G1M1Lum]{Fig. \ref*{fig:G1M1Lum}} shows the luminosity of each radiative group (computed using \hyperref[eq:lumr]{Eq. \ref*{eq:lumr}}). We see that the total multigroup luminosity (lime dash-dotted line) and the gray luminosity (black dotted line) are very similar, albeit the location of the first core accretion shock differs slightly (0.5 and 0.6 $\rm{AU}$). This is due to a slightly higher amount of enclosed radiative energy inside the first core for the multigroup run (in turn due to a higher opacity for UV-Visible photons), which causes its specific entropy to increase by virtue of radiative heating from the second core accretion shock.
\\
Unsurprisingly, the UV-Visible group dominates the luminosity output of the protostar, whereas the IR group dominates everywhere else. At both first and second core shock fronts, the luminosity of each radiative group spikes, although the X-ray photons produced at these locations are quickly reprocessed by the other groups.
\\
\\
Finally, the evolution of the properties of the protostar formed in the multigroup run is compared to that of its gray counterpart in \hyperref[fig:G1M1Evol]{Fig. \ref*{fig:G1M1Evol}}. We find that the radii, luminosities and radiative efficiencies are extremely similar, although the mass differs slightly. As mentioned previously, the first core in the multigroup run has a slightly higher enclosed energy. This causes the mass accretion rate unto the second core to be lower. Despite the differing masses, the radii are very similar because of a similar amount of specific entropy.
\\
\\
This allows us to conclude that the multigroup description offers no major differences to its gray counterpart, a result which is in agreement with \cite{vaytet_2013}.

\begin{table}[]
    \centering
    \begin{tabular}{| c | c | c |}
        \hline
          Radiative Group & [$\nu_1 ; \nu_2$] ($\rm{Hz}$) & [$\lambda_2 ; \lambda_1$] ($\rm{m}$)\\
         \hline
         1: Radio & [$10^{5} ; 3\times10^{11}$] & [$3\times10^{3};10^{-4}$]\\
         \hline
         2: IR & [$3\times10^{11} ; 4.287\times10^{14}$] & [$10^{-4};7\times10^{-7}$]\\
         \hline
         3: UV-Visible & [$4.287\times10^{14} ; 3\times10^{16}$] & [$7\times10^{-7};10^{-8}$]\\
         \hline
         4: X-ray & [$3\times10^{16} ; 10^{19}$] & [$10^{-8};3\times10^{-11}$]\\
         \hline
    \end{tabular}
    \caption{Frequency and corresponding wavelength borders of the 4 radiative groups used in our multigroup simulation.}
    \label{tab:radGroups}
\end{table}

\section{Comparison with a 2D simulation} \label{appendix:bhandareComparison}
\begin{figure}
    \centering
    \includegraphics[width=.45\textwidth]{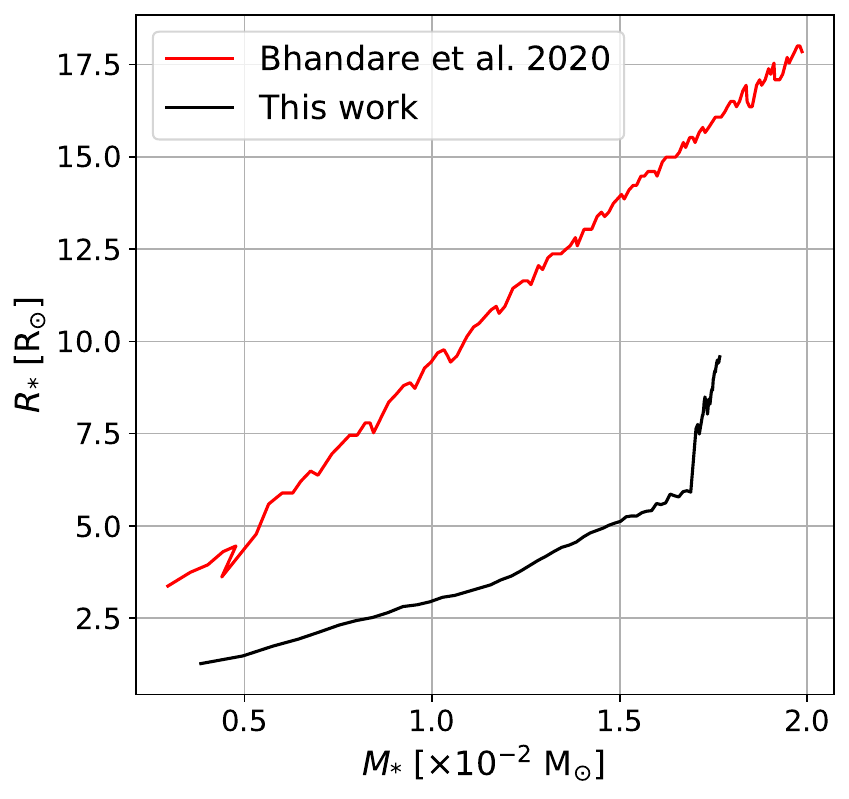}
    \caption{Radius of the protostar as a function of its mass: a comparison of the results of this paper (black curve) with those of \cite{bhandare_2020} (red curve).}
    \label{fig:rstarmstar2D}
\end{figure}

\begin{figure}
    \centering
    \includegraphics[width=.45\textwidth]{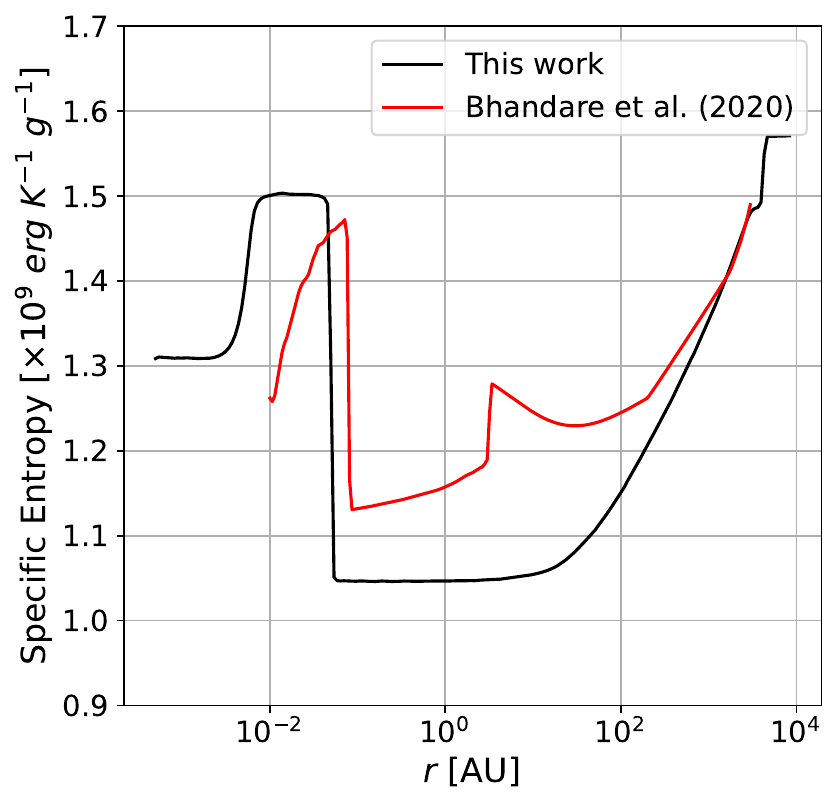}
    \caption{Comparison of the results of this paper (black curve) with those of \cite{bhandare_2020} (red curve). The curves display the specific entropy, averaged in radial bins and displayed as a function of radius, at a moment in time where both protostars have a mass of $\approx 1.76\times 10^{-2}\ \mathrm{M_{\odot}}$.}
    \label{fig:entropy2D}
\end{figure}

Herein, we compare the results of our simulation with those of the $1\ \mathrm{M_{\odot}}$ 2D collapse calculations of \cite{bhandare_2020}. Since their calculations are similar to ours, this will allow us to better assess what a three dimensional description of the gas motion offers. To this end, we begin by studying \hyperref[fig:rstarmstar2D]{Fig. \ref*{fig:rstarmstar2D}}, which shows the protostellar radius as a function of protostellar mass. We see that protostar is consistently more compact than in the 2D calculation; it possesses a smaller radius for a given mass. Since the radiative efficiency of the protostar is extremely low in both simulations, this cannot be explained by any of the protostars radiating away more energy than the other. We explain this difference in radii by comparing their radial entropy profiles in \hyperref[fig:entropy2D]{Fig. \ref*{fig:entropy2D}}: by the time the protostar reaches a mass of $\approx 1.76\times 10^{-2}\ \mathrm{M_{\odot}}$, the entropy plateau in the interior is achieved in our 3D simulation, whereas it has yet to form in the 2D counterpart. This allows us to conclude that the 3D description of the gas motion allows for more efficient entropy mixing, which regulates the radius of the protostar.
\\
However, the entropy profile outside the second core is quite different in our two simulations. This can be explained by the different initial conditions. Indeed,  \cite{bhandare_2020} have used a Bonnor-Ebert sphere as their initial conditions, whereas we have used a highly unstable uniform density sphere. This results in a shorter first core lifetime in our simulation, and it is accreted by the time our protostar has reached $\approx 1.76\ \mathrm{M_{\odot}}$. In addition to this, the equation of state table used in both simulations is different. This causes different behaviors in entropy, particularly inside the second Larson core since the \cite{saumon_1995} EOS takes into account the ionization of He, whereas the \cite{vaidya_2015} EOS used in \cite{bhandare_2020} does not.

\section{Standing Accretion Shock Instability}\label{appendix:sasi}
\begin{figure}
    \centering
    \includegraphics[width=.45\textwidth]{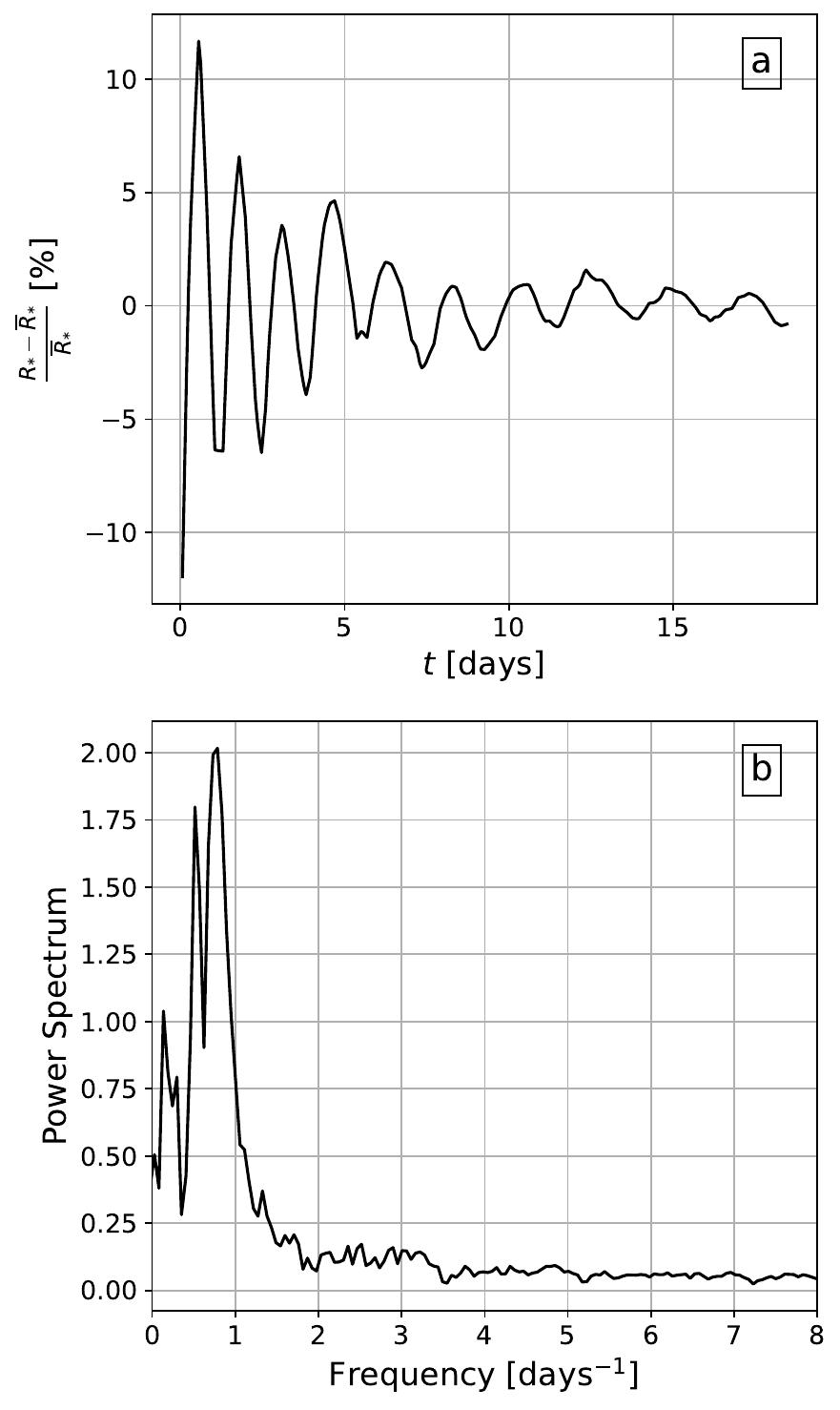}
    \caption{Amplitude of oscillations of the protostellar radius in the $\ell_{\mathrm{max}}=27$ run (panel a), displayed as a function of time where $t=0$ marks the birth of the protostar. Panel (b) displays the Fourier transform of the curve in panel (a).}
    \label{fig:oscillations}
\end{figure}

Herein, we investigate wether or not the Standing Accretion Shock Instability (SASI, \citealp{blondin_2003, sheck_2004, foglizzo_2007}) could be the mechanism behind the onset of turbulence within our protostar. This instability is known to operate in core-collapse supernovae, where it causes them to appear aspherical. Recently, \cite{bhandare_2020} put forth the hypothesis that this instability could be at play in protostars, where it could cause oscillations of the accretion shock. SASI requires for feedback to occur between the central regions and the accretion shock. Although our physical environment heavily differs from that of a core-collapse supernova and we do not have a proto-neutron star downstream of our second core accretion shock, our protostar has a central region of highly dense, ionized gas that repulses inward flow. In this sense, the central regions of our protostar can communicate with the accretion shock through acoustic feedback. In the 2D study of \cite{bhandare_2020}, the central regions ($r<10^{-2}\ \mathrm{AU}$) were a part of a reflexive inner boundary, which can naturally provide feedback to the shock front.
\\
\\
In order to investigate whether this mechanism is responsible for the generation of turbulence in our protostar, we study our $\ell_{\mathrm{max}}=27$ run presented in \hyperref[appendix:ResStudy]{Appendix \ref*{appendix:ResStudy}} due to its high spatial and temporal resolution. Indeed, this run presents oscillations of the protostellar radius possibly caused by SASI. To this end, we display in panel (a) of \hyperref[fig:oscillations]{Fig. \ref*{fig:oscillations}} the amplitude of said oscillations, computed as $(R_{*}-\overline{R}_{*})/\overline{R}_{*}$, where $\overline{R}_{*}$ is the average radius of the protostar over a given period. Here, we can clearly see high amplitude, high frequency oscillations at protostellar birth; however, their amplitude and frequency reduces over time. This is more readily seen in the power spectrum of this curve (panel b), which shows a high energy peak corresponding to a period of $\approx$ 1.4 days, and a handful of lower energy low frequency peaks. These oscillation periods of the protostellar radius should be compared with the advection timescale $t_{\mathrm{adv}}$, computed as \citep{foglizzo_2007}:
\begin{equation}
    t_{\mathrm{adv}} = \int_{R_{\mathrm{\nabla}}}^{R_{*}}\frac{dr}{|v_{\mathrm{r}}(r)|},
\end{equation}
where $R_{\mathrm{\nabla}}$ is the radius where the gas has effectively settled following its crossing of the accretion shock (i.e., $v_{\mathrm{r}}$ has reached $\approx 0$). Our estimate of $t_{\mathrm{adv}}$ has yielded $\approx$ 3 days, which is about twice as long as the oscillation period of the protostar. However, as the protostellar radius grows, so too does $t_{\mathrm{adv}}$, which could explain why the frequency of oscillations is reducing over time.
\\
\\
Although these measurements do not allow us to conclude with certainty that SASI is operating in our protostar, they do indicate that we are in the regime where it is theoretically possible.

\end{appendix}
\end{document}